\documentclass[epjST]{svjour}
\usepackage{mathrsfs}
\usepackage{amsmath}
\usepackage{amssymb}
\usepackage{hyperref}
\usepackage{graphics}
\usepackage{epsfig}
\usepackage{rotating}
\usepackage{epstopdf}
\usepackage{cite}
\usepackage{color}
\usepackage{ulem} 

\newcommand{\mL}{\mathcal{L}}
\newcommand{\mT}{\mathcal{T}}
\newcommand{\bra}{\langle}
\newcommand{\ket}{\rangle}
\newcommand{\nn}{\nonumber}
\newcommand{\kv}{K^*(892)}
\newcommand{\rhov}{\rho(770)}

\begin{document}
\title{$N_C$ evolution of the light meson resonances}
%


\author{Zhi-Hui Guo \inst{1\,,}\inst{2}\fnmsep\thanks{\email{zhguo@seu.edu.cn}}  }
\institute{  School of Physics, Southeast University, Nanjing 211189, China
\and Department of Physics and Hebei Advanced Thin Films Laboratory, Hebei Normal University,  Shijiazhuang 050024, China;  }

\abstract{ 
The variation of the resonance poles, i.e., the mass and width, originating from changing the values of $N_C$, the number of the QCD colors, provide an intuitive theoretical method to discriminate the interior constituents of hadrons. Many different approaches, including unitarized chiral amplitudes, dispersive methods, quark models, linear-sigma-model-like framework, etc., have been widely used to determine the $N_C$ trajectories of the various resonance pole positions.  We focus on the discussions of the light-flavor meson resonances in this work.  
} 
\maketitle
\section{Introduction}
\label{intro}

Discerning the internal structures of the hadron resonances is a highly nontrivial task, due to the quark confinement feature and the nonperturbative nature of the QCD in the resonance energy region. Sophisticated theoretical methods, based on the general principles of the S matrix theory (e.g., unitarity and analyticity) and  chiral perturbation theory ($\chi$PT), have been impressively developed over the last decades to improve the descriptions of the various data from the experiments and lattice QCD~\cite{Pelaez:2015qba,Yao:2020bxx,Oller:2019opk,Oller:2020guq}. The hadronic states are typically included as point-like particles~\footnote{The reference to point-like particles here means that they are not treated as compounds with explicit smaller constituent parts as compared to e.g., a quark-model description.}, in the effective field theory approach, from which alone it is difficult to get information about the interior structures of the hadrons. Consequently, additional theoretical ingredients are usually needed to further discriminate the possible constituents inside the various hadrons. One of such methods is the compositeness relation developed by Weinberg in the study of the hadronic bound-state deuteron~\cite{Weinberg:1962hj}, where the probability to find the two-nucleon state in the deuteron, i.e., the compositeness coefficient, is calculated. There exist several recipes to generalize the compositeness relation to the resonance case from several  groups~\cite{Baru:2003qq,Hanhart:2011jz,Hyodo:2011qc,Aceti:2012dd,Sekihara:2014kya,Guo:2015daa,Oller:2017alp,Matuschek:2020gqe}. The merit of the compositeness relation approach is that one can give quantitative estimation of the probabilities to find specific  constituting bodies in the resonances. However, up to date, a general agreement for the generalization of the compositeness to the resonance is not reached yet~\cite{Gao:2018jhk} and different recipes could give quite different results for the compositeness coefficients. Other qualitative methods to discriminate different constituent configurations of the various hadrons include the changing of the QCD parameters, such as the number of colors and the quark masses. Hadrons with different constituent configurations are expected to exhibit different responses when changing these parameters. The $\chi$PT naturally provides a reliable theoretical framework to investigate the quark-mass dependences of the hadron properties~\cite{Hanhart:2008mx,Hanhart:2014ssa,Liu:2012zya,Torres:2014vna,Guo:2015dha,RuizdeElvira:2017aet,Ren:2012aj}, which can be further  examined by the lattice calculations~\cite{Aoki:2019cca}.

Another powerful tool to study the possible interior structures of the hadron resonances is the large $N_C$ QCD~\cite{largenc}, being $N_C$ the number of the colors. By taking the scaling of the strong coupling of QCD as $g_s^2/N_C$ when $N_C\to\infty$ and assuming the confinement, the leading $N_C$ scaling of the masses and widths for the conventional meson composed of a quark and an anti-quark ($\bar{q}q$) and the glueball composed of pure gluons can be unambiguously determined~\cite{largenc}. E.g., the large $N_C$ QCD predicts that the masses of both types of hadrons approach to constants, and the widths of the $\bar{q}q$ meson and the glueball scale as $1/N_C$ and $1/N_C^2$, respectively. Despite these leading $N_C$ scaling laws are obtained in the large $N_C$ QCD, they should hold with corrections at finite values of $N_C$.

For the exotic mesons beyond the conventional $\bar{q}q$ picture, specially those composed of two quarks and two antiquarks, the situation is more subtle. It has long been thought that the tetraquark meson state is not allowed by the large $N_C$ QCD~\cite{Coleman:1985rnk}. The main reason was that the leading-order (LO) contributions of the correlation functions of the two tetraquark  operators are the disconnected parts, containing two closed quark loops, which correspond to the two color-singlet mesonic states at the hadronic level. Therefore it was concluded in Ref.~\cite{Coleman:1985rnk} that the tetraquark meson does not exist in the large $N_C$ QCD. However, in 2013 the long standing view of the absence of tetraquark states was questioned in Ref.~\cite{Weinberg:2013cfa} and many other following works~\cite{Knecht:2013yqa,Lebed:2013aka,Cohen:2014tga,Cohen:2014vta,Maiani:2016hxw,Lucha:2017gqq}. The key argument is that the subleading order connected parts of the two-tetraquark-operator correlators, although $1/N_C$ suppressed comparing with the leading disconnected contributions, could contain the possible tetraquark meson poles, whose decay widths in the large $N_C$ limit can be proportional to $1/N_C$ or even narrower ones for specific flavor contents~\cite{Weinberg:2013cfa,Knecht:2013yqa,Lebed:2013aka,Cohen:2014tga,Cohen:2014vta,Maiani:2016hxw,Lucha:2017gqq}. In this way, the tetraquark mesons can have narrow widths and may be relevant for various physical processes. 

A much related type of hadron to the tetraquark state is the two-meson molecule. A well accepted rule to distinguish the genuine tetraquark and two-meson molecule is far to be reached. From the large $N_C$ point of view, the hadron molecule made of two color-singlet objects may be likely to have a large width when increasing $N_C$. Indeed in the large $N_C$ the extraordinary scaling laws  for the widths of the molecule-like resonances, such as those with $O(1)$, $O(N_C)$ or even higher powers of $N_C$, are not found to be incompatible with the known properties of the dispersion relations and effective theories~\cite{Cohen:2014vta,Jaffe:2007id,Jaffe:2008zz}. In the $N_C>3$ case, there could exist another type of interesting hadronic states, composed of $(N_C-1)$ quarks and $(N_C-1)$ antiquarks, whose mass can behave as $O(N_C)$ and width as $O(1)$~\cite{Jaffe:1981,Jaffe:2007id,Cohen:2014vta,Cohen:2014tga}. Similar large $N_C$ discussions can be also extended to the baryons, whose masses and widths will scale as $N_C$ and constants, respectively~\cite{Dashen:1993jt,Manohar:1998xv,Goity:1996hk,Lutz:2001yb}. In this short note, the topics on the baryons at large $N_C$ will not be covered and we focus on the $N_C$ properties of the meson resonances, specially concentrating on the light-flavor meson resonances, and the related physical quantities.

The classification of the various hadrons, according to the scaling laws of their masses and widths discussed above, is inferred from the large $N_C$ QCD. In many cases, we are more interested in the physical hadrons living in the real world with $N_C=3$. As a result, the $N_C$ trajectories of the hadron properties obtained from $N_C=3$ to larger values of $N_C$, or even to $N_C\to \infty$, can provide important information to discern the internal structures of the physical resonances observed in experiments at $N_C=3$. Nevertheless to extrapolate the hadron properties to $N_C>3$, especially to rather large values of $N_C$, is not an easy task and in many cases it bears big uncertainties. Furthermore, it is quite possible that the physical hadrons are complicated mixtures of different types of constituents with different $N_C$ behaviors. In the ideal case when the physical resonance has little mixing of constituents with different $N_C$ behaviors, the $N_C$ trajectory of this resonance pole is expected to exhibit simple scaling laws of the dominant constituent. E.g., the vector resonances, $\rho$ and $\kv$, belong to such category of resonances. The $N_C$ trajectories of their resonance poles from various studies, as shown in later discussions, follow almost exactly as that of the $q\bar{q}$ state, with their masses behaving as constants and widths as $1/N_C$ in a wide range from $N_C=3$ to rather large values up to one hundred or even bigger ones. While in the complex situation when the physical hadron resonance is a mixture of several types of constituents with different $N_C$ behaviors, the $N_C$ trajectory becomes complicated and it may show different trends in different $N_C$ regions. An even more subtle situation is that if at $N_C=3$ the hadron is dominated by a constituent which is however suppressed at large $N_C$, its trajectories in the region of large values of $N_C$  likely contain big uncertainties. Since the hadron properties at $N_C=3$ are constrained by the relevant experimental data, the theoretical uncertainties in the extrapolation to the nearby values around $N_C\ge 3$ are believed to be under better control than the situation with rather large values of $N_C$~\cite{Pelaez:2015qba,Pelaez:2010er}. Therefore it is advisory to separately address the results in the regions near $N_C=3$ and those in large values of $N_C$. With this in mind we address in this review the evolution in $N_C$ of the masses and widths of various light meson resonances and related physical quantities.

\section{State-of-art study of the $N_C$ trajectories for light meson resonances }
\label{sec:2}

To keep track of the $N_C$ movement of the hadron resonance is by no means a trivial matter, since one has to first determine the resonance properties from the experimental data, corresponding to the initial point of the $N_C$ trajectory at $N_C=3$, and then to properly perform the extrapolations to larger values of $N_C$. One of the practicable and reliable ways to investigate the $N_C$ evolution of the hadron resonance is using unitarized $\chi$PT in different forms, with Ref.~\cite{Oller:1998zr} being the pioneer one to signal out the remarkable unusual scaling of the poles of the lightest scalar resonances in powers of 1/$N_C$ within that framework.  A heavily used approach along these lines is the inverse amplitude method (IAM)~\cite{Pelaez:2003dy}, which relies on the combination of the $\chi$PT and the requirements of the general properties of the scattering amplitudes, such as the unitarity and analyticity. To set up the notations, we briefly recapitulate the methodology how the $N_C$ trajectories of the resonance poles are obtained in the IAM approach. 

Chiral perturbation theory, the low energy effective realization of QCD, takes the pseudo Nambu-Goldstone bosons (pNGBs), e.g., $\pi$ in the $SU(2)$ case and $\pi,K,\eta$ in the $SU(3)$ version, as the dynamical fields. By construction $\chi$PT respects the important spontaneous and explicit chiral symmetry breaking patterns of QCD, and it has a rigorously organized power counting rule to calculate the physical quantities order by order in perturbation theory. The chiral power counting rule is based on the perturbative expansion of the external momentum $p$ and the light quark masses $m_q$, counted as $O(p^2)$~\cite{Weinberg:1978kz,Gasser:1983yg,Gasser:1984gg}. 

At leading order in the $SU(3)$ $\chi$PT, all the free parameters in the Lagrangian can be fixed by the pion weak decay constant in the chiral limit, and the pion and kaon masses. In fact all the decay constants of the pNGBs, denoted as $F$, share the same value at leading order. The mass of the $\eta$ at this order can be completely predicted, and the result coincides with the celebrated Gell-Mann-Okubo relation $m_\eta^2=(4m_K^2-m_\pi^2)/3$. At next-to-leading order (NLO) of the $SU(3)$ $\chi$PT, twelve additional $O(p^4)$ local operators, each accompanied by one low energy constant (LEC), will appear and eight of them are relevant to the meson-meson scattering. The relevant LECs are labeled as $L_{i=1,2,\cdots,8}$~\cite{Gasser:1984gg}. For the perturbative NLO meson-meson scattering amplitudes, they include the tree-level and one-loop diagrams. All the divergences of the one-loop diagrams can be completely canceled by the infinite parts of the $O(p^4)$ LECs. The remaining unknown finite parts of the LECs, which are usually needed to be fitted to the experimental or lattice data, together  with the finite parts of the one-loop diagrams, give the final expression for a physical quantity. To confront the $\chi$PT with the large $N_C$ QCD, e.g. comparing the various Green functions calculated in both schemes, enables one to obtain the leading $N_C$ scaling of the parameters from $\chi$PT~\cite{Gasser:1984gg}. The results are summarized as
\begin{eqnarray}\label{eq.nclecs}
&& F^2\sim O(N_C)\,,  \quad  L_{1}\sim O(N_C)\,, \quad  L_{2}\sim O(N_C)\,, 
\quad 2L_{1}-L_2\sim O(1)\,, \quad L_{3}\sim O(N_C)\,, \nonumber \\ && 
 L_{4}\sim O(1), \quad L_{5}\sim O(N_C), \quad
L_{6}\sim O(1), \quad L_{7}\sim O(1), \quad L_{8}\sim O(N_C)\,. 
\end{eqnarray}
The $N_C$ scaling of $L_7$ written above is given by counting the number of traces. In fact the $N_C$ counting of $L_7$ in the $SU(3)$ $\chi$PT is more subtle, due to the contribution from the heavy singlet $\eta_0$~\cite{Gasser:1984gg,Peris:1994dh}, whose mass squared scales as $1/N_C$. One of the proper ways to address this problem is the $U(3)$ $\chi$PT, which will be discussed in detail later. Notice that it is possible that some of the LECs may contain sizable subleading $N_C$ correction terms~\cite{Ledwig:2014cla}. 
And the masses of the pNGBs behave as constants at large $N_C$, i.e.  $m_{\pi,K,\eta}^2\sim O(1)$. 

The perturbative amplitudes from the $\chi$PT calculation alone can not generate any resonance state. The requirement of the unitarity for the partial-wave scattering amplitude provides a crucial guideline to extend the perturbative $\chi$PT to the resonance energy region. The IAM is one of such attempts to build unitarized partial-wave amplitudes and its explicit expression for the two-meson scattering processes with definite isospin ($I$) and angular momentum ($J$) takes the form 
\begin{eqnarray}\label{eq.iam}
 \mT_{IJ}(s)= T_{2,IJ}(s)\cdot \big[T_{2,IJ}(s)-T_{4,IJ}(s)\big]^{-1}\cdot T_{2,IJ}(s) \,,
\end{eqnarray}
where the subscripts $2$ and $4$ denote the chiral orders of the perturbative amplitudes. In the couple-channel scattering, $T_{2,IJ}(s)$ and $T_{4,IJ}(s)$ in the above equations should be understood as matrices spanned in the scattering-channel space. For simplicity, we will often omit the subscripts $IJ$ when it is not necessary to clarify them. The partial-wave amplitude $T_2(s)$ corresponds to the tree-level results calculated from the LO Lagrangian, which are completely fixed by the weak decay constants of the pNGBs and their masses. For the $O(p^4)$ amplitude $T_4(s)$, it receives contributions both from the one-loop diagrams calculated with the LO Lagrangian and the tree-level diagrams from the NLO Lagrangian. The simple IAM amplitude has been demonstrated to be quite powerful to describe the various experimental scattering data~\cite{Oller:1997ng,Oller:1998hw,GomezNicola:2001as}. After the unknown LECs are determined from the fits to data, one can extrapolate the scattering amplitudes into the complex energy plane to look for the resonance poles, the real and imaginary parts of which are usually identified as the masses and half widths of the resonances, respectively. For the IAM in the $SU(3)$ case, the scalar resonances $f_0(500)/\sigma$, $K_0^*(700)$, $f_0(980)$ and $a_0(980)$ and the vector resonances $\rho(770)$ and $K^*(892)$ can be naturally obtained~\cite{Oller:1997ng,Oller:1998hw,GomezNicola:2001as}. From the $N_C$ point of view, the LECs determined from the fits, correspond to their values at $N_C=3$. Though the resulting values of the LECs from the unitarized $\chi$PT should be qualitatively comparable to those from the standard $\chi$PT~\cite{Gasser:1983yg,Bijnens:2014lea}, moderate variations between the two approaches are also within expectations, because the LECs might receive noticeable  contributions from higher orders obtained by iterating procedure. Nevertheless, the values of the LECs from the standard $\chi$PT provide a useful guide for the unitarized-$\chi$PT fits. It is advisory to rely on the reasonable unitarized fits, that lead to comparable values of LECs to the standard $\chi$PT, to  perform the $N_C$ studies by imposing the leading $N_C$ scaling shown in Eq.~\eqref{eq.nclecs}.

The resonance poles from the IAM amplitudes are determined by the LECs, such as the $F$ and $L_{i=1,2,\cdots,8}$. Since the $N_C$ scalings of the various LECs are known~\eqref{eq.nclecs}, one can implement them to get the LECs at different values of $N_C$ through 
\begin{eqnarray}\label{eq.linc1}
L_i(N_C)= L_i^{\rm Exp} \frac{N_C}{3}\,,  
\end{eqnarray}
for $L_i\sim O(N_C)$ and 
\begin{eqnarray} \label{eq.linc0}
L_i(N_C)= L_i^{\rm Exp}\,,
\end{eqnarray}
for $L_i\sim O(1)$, where $L_i^{\rm Exp}$ denote the values obtained from the phenomenological fits to the experimental data. The decay constant $F$ will be scaled as 
\begin{eqnarray} \label{eq.lincfpi}
 F(N_C)= F^{\rm Exp}\sqrt{\frac{N_C}{3}}\,.
\end{eqnarray}
By taking into account the results of the LECs obtained at different values of $N_C$, one can then update the calculations of the resonance poles correspondingly. In such a way, the paths of the resonance poles at different values of $N_C$ can be tracked. Slight different strategies are adopted by several research groups to address the $N_C$ trajectories of the various light-flavor meson resonances. Below we highlight the main findings from different groups.

\subsection{Resonance trajectories from the IAM with $SU(3)$ $\chi PT$ at one loop}
\label{sec:21}

The first attempt to the investigate the $N_C$ evolution of the resonance poles in the IAM approach is carried out in Refs.~\cite{Pelaez:2003dy,Pelaez:2004xp}, which is based on the one-loop two-meson scattering amplitudes of the $SU(3)$ $\chi$PT~\cite{GomezNicola:2001as}. The coupled-channel IAM formalism is used to fit many experimental scattering data, including the phase shifts and inelasticities from the $\pi\pi\to\pi\pi$, $\pi\pi\to K\bar{K}$, $\pi K\to\pi K$ in different partial waves and isospin channels, and also the $\pi\eta$ event distributions~\cite{GomezNicola:2001as}. For the isoscalar-scalar ($IJ=00$) case, three couple channels, i.e. $\pi\pi, K\bar{K}, \eta\eta$, are included in the IAM amplitudes, and the two scalar resonances $\sigma$ and $f_0(980)$ are found to be compatible with nowadays determinations~\cite{Tanabashi:2018oca}. For the $(I,J)=(1,1)$ case, $\pi\pi$ and $K\bar{K}$ channels are incorporated, and the vector resonance $\rho(770)$ found in the complex energy plane agrees well with the PDG~\cite{Tanabashi:2018oca}. For the $(I,J)=(1/2,1)$ and $(1/2,0)$ cases, the $K\pi$ and $K\eta$ channels are included. The vector $K^*(892)$ resonance is nicely reproduced and a clear resonance pole for the scalar $K^*_0(700)/\kappa$ is confirmed. For the $(I,J)=(1,0)$ case, two couple channels $\pi\eta$ and $K\bar{K}$ are taken into account, and a resonant pole is found for the $a_0(980)$. Other experimental scattering data in the elastic-channel cases, including the $\pi\pi$ with $(I,J)=(2,0)$ and $K\pi$ with $(I,J)=(3/2,0)$, are also used to constrain the $O(p^4)$ LECs, although resonances do not appear in these two cases.

To impose the leading $N_C$ scalings for various LECs in Eq.~\eqref{eq.nclecs}, the movements of the resonance pole positions by varying the values of $N_C$ are illustrated in Refs.~\cite{Pelaez:2003dy,Pelaez:2004xp}. The masses of the vector resonances $\rho(770)$ and $K^*(892)$ are barely changed when increasing the values of $N_C$ up to 30, while their widths clearly decrease and nicely exhibit the $1/N_C$ scaling law. Those $N_C$ behaviors of the vector resonances are well compatible with the expectations for the $\bar{q}q$ meson resonances. The results in the vector channels also provide solid evidences that the resulting $N_C$ trajectories of the resonances from the IAM study do make sense, since there is little doubt that the $\rhov$ and $\kv$ are the well established hadrons overwhelmingly dominated by the $\bar{q}q$ components~\cite{Tanabashi:2018oca}. In contrast, the $N_C$ evolutions for the scalar resonances $\sigma$, $f_0(980)$, $\kappa$ and $a_0(980)$ are rather different from the vectors. When increasing the values of $N_C$ up to 30, it is found that the masses of the $\sigma$, together with the uncertainties, tend to increase. With moderate uncertainties, the masses of the $\kappa$ are also found to increase. The widths of the $\sigma$ and $\kappa$ both steadily keep increasing at the rates between $\sqrt{N_C}$ and $N_C$. Based on these observations, it was concluded that the $\sigma$ and $\kappa$ resonances are likely to be dominated by the tetraquark or two-meson components, instead of the $\bar{q}q$  constituents~\cite{Pelaez:2003dy,Pelaez:2004xp}.

In a later study~\cite{Sun:2005uk}, similar $\sigma$ trajectories are obtained as those in Refs.~\cite{Pelaez:2003dy,Pelaez:2004xp} in the region of  $N_C\lesssim 30$. By further taking very large values of $N_C$ in Ref.~\cite{Sun:2005uk}, say several hundreds or more, the $\sigma$ poles are found to fall down to the positive or negative real axis in the complex $s$ plane, depending on the values of the LECs. This shows that for some of the LECs the decay width of the $\sigma$ could vanish for such huge $N_C$, indicating that a $q\bar{q}$ seed could exist in the formation of the $\sigma$. Therefore it is concluded in Ref.~\cite{Sun:2005uk} that the $\sigma$ seems a $\bar{q}q$ resonance surrounded with important $\pi\pi$ clouds. The rather different $\sigma$ pole trajectories in the region around $N_C=3$, as compared to the $\rho$, imply that the $\pi\pi$ or tetraquark components seem playing much more important roles than the case of $\rho$ for $N_C$ not far away from 3. As mentioned in the Introduction and discussed in detail in Refs.~\cite{Pelaez:2015qba,Pelaez:2010er}, it could bring in big theoretical uncertainties when taking rather large values of $N_C$ to determine the resonance properties. E.g., it is estimated that the 10-20\% error bars at $N_C=3$ in the unitarized $\chi$PT~\cite{Pelaez:2010er,Salas-Bernardez:2020hua} can be amplified to 100\% errors when taking $N_C=30$ or 15. Also a slight change of the values of the LECs could result in rather different behaviors of the pole trajectories at large $N_C$, as discussed in Ref.~\cite{Sun:2005uk}. Although the resonance trajectories at large $N_C$ may reveal some components that are not the dominant ones at $N_C=3$, one has to first overcome the somewhat big uncontrollable theoretical uncertainties to reach more definite conclusions in the large $N_C$ region.

Due to the closeness of the $K\bar{K}$ threshold, the $N_C$ behaviors of the $f_0(980)$ and $a_0(980)$ are more complicated than those of the $\sigma$ and $\kappa$. Their poles in the complex energy plane were found to be difficult to track for larger values of $N_C$~\cite{Pelaez:2004xp}. According to the $N_C$ evolution of the scattering amplitudes, the $f_0(980)$ signal can be peaks, dips, or even disappear for different values of $N_C$. The scattering amplitudes relevant for the $a_0(980)$ resonances turn out to be sensitive to the regularization scale $\mu$. When taking $N_C=25$, there can be a peak around the $a_0(980)$ energy region for $\mu=550$~MeV, or a rather smooth amplitude for $\mu=770$~MeV. The $N_C$ behaviors of the amplitudes relevant for the $f_0(980)$ and $a_0(980)$ show that most likely the $\bar{q}q$ components play small roles in these two scalar resonances~\cite{Pelaez:2003dy,Pelaez:2004xp}. Nevertheless, it is still difficult to reach definite conclusions about what the dominant constituents inside the $f_0(980)$ and $a_0(980)$ are.

\subsection{ Shadow poles in the [1,1] Pad\'e of the $SU(3)$ $\chi PT$}  

In Refs.~\cite{Dai:2011bs,Dai:2012kf}, the authors propose to use the pole counting rule~\cite{Morgan:1992ge} to analyze the numbers of resonance poles at different values of $N_C$. The one-loop IAM or the [1,1] Pad\'e amplitudes based on the one-loop $SU(3)$ $\chi$PT are used to revise the analyses of the resonance poles, especially emphasizing the relevant shadow poles in different Riemann sheets (RSs) for the resonances. According to the pole counting rule, an elementary resonance with dominant $\bar{q}q$ component is characterized by the presence of all the relevant shadow poles in different RSs. In contrast, for the molecular type of hadron resonance, usually there is only one resonant pole. The shadow poles in the all the relevant RSs for the vector resonances of $\rhov$ and $\kv$ can be easily tracked for different values of $N_C$, which reconfirm their $\bar{q}q$ nature. Definite conclusion is not obtained for the $a_0(980)$ due to the large uncertainties of the data~\cite{Dai:2012kf}. No relevant shadow poles are found for $f_0(980)$ when varying the values of $N_C$, and it supports the $K\bar{K}$-molecule explanation of the $f_0(980)$. For the $\sigma$ and $\kappa$, although the shadow poles in different RSs are found at $N_C=3$ and also for larger values of $N_C$, the poles in different RSs turn out to be located in somewhat different positions~\cite{Dai:2011bs,Dai:2012kf}. It can be concluded that the $\sigma$ and $\kappa$ only roughly meet the ``Breit-Wigner criteria''~\cite{Dai:2011bs,Dai:2012kf}~\footnote{I would like to thank Ling-Yun Dai for useful discussions on these two references.}. The poles of $\sigma$ and $\kappa$ tend to drive away from the real $s$ axis when increasing the values of $N_C$ up to moderate values not far away from three, which are clearly different from the $N_C$ trajectories of the well established $q\bar{q}$ resonances $\rho$ and $\kv$. This indicates that the $q\bar{q}$ seeds are unlikely the dominant components in the $\sigma$ and $\kappa$. Furthermore, the poles of the $\sigma$ and $\kappa$ are found to fall down to the negative real axis in the $s$ plane when taking huge values of $N_C$, indicating that they become irrelevant at large $N_C$.

\subsection{Fates of the resonance poles from the $SU(2)$ two-loop IAM} 

The sensible results obtained in the one-loop IAM study are further verified in the two-loop IAM formalism~\cite{Sun:2005uk,Pelaez:2006nj}. Up to now the IAM study at two-loop order has been only extended to the $SU(2)$ $\chi$PT, in which the additional $O(p^6)$ LECs relevant to the two-meson scattering are still manageable~\cite{Bijnens:1995yn}. The $SU(2)$ two-loop IAM quantitatively confirms the $N_C$ behaviors for the $\rhov$ from the $SU(3)$ one-loop study~\cite{Pelaez:2006nj}. While for the $\sigma$, it is found that a $\bar{q}q$ component with the mass around $1$~GeV could play relevant roles in its formation, although the seed of scalar state around 1~GeV is unlikely to be important for the $\sigma$ at $N_C=3$. When increasing the values of $N_C$, the two-meson loops are further suppressed and the $\bar{q}q$ seed becomes more important, which drives in the large values of $N_C$ the $\sigma$ pole down to the real axis around 1~GeV~\cite{Pelaez:2006nj,Sun:2005uk}, rather different from its values at $N_C=3$. Nevertheless it is found that the $\bar{q}q$ source at $1$~GeV in the $\sigma$ resonance plays quite important roles to fulfill the semilocal duality~\cite{RuizdeElvira:2010cs}. It is also further verified in Ref.~\cite{Pelaez:2006nj} that with reasonable NNLO parameters it is impossible to make the $\sigma$ scale as a conventional $q\bar{q}$ meson with  an $N_C$ dependent mass close to the its physical mass at $N_C=3$, without destroying the well established $N_C$ behaviors of the $\rho$ resonance. 

While the $N_C$ trajectories of the $\rhov$ are quite robust, the $N_C$ paths of the $\sigma$ pole are subject to many ingredients, such as the subleading correction of the $N_C$ scaling of the LECs, higher order amplitudes, slight different fit strategies and so on. The large uncertainties of the $\sigma$ trajectories with varying $N_C$ can be easily seen in Fig.~2 of Ref.~\cite{RuizdeElvira:2010cs}. By tuning the regularization scale $\mu$ from 500~MeV to 1000~MeV, which can be considered as one part of the subleading $N_C$ scaling corrections, rather different fates for the $\sigma$ poles show up. For $\mu=500$~MeV and $1000$~MeV, the $\sigma$ poles tend to move to the negative and positive real axis in the complex energy squared $s$ plane, respectively. For $\mu=770$, the $\sigma$ poles seem running deeply in the complex plane for rather large values of $N_C$. A deep  explanation for the large uncertainties of the $\sigma$ trajectories is obtained by studying the analytical pole positions in the chiral and large $N_C$ limit~\cite{Sun:2005uk,Nieves:2009ez,Cohen:2014vta}, which reveals a crucial combination of the chiral LECs, i.e. $25L_2+11L_3$, that will determine the fates of the $\sigma$ in the large $N_C$ limit. Intriguingly the phenomenologically preferred values of the chiral LECs almost exactly lead to the critical value for the combination~\cite{Sun:2005uk,Nieves:2009ez,Cohen:2014vta}. Unless the chiral LECs and their subleading $N_C$ scalings can be precisely determined, more definite conclusions on the $\sigma$ trajectories for large values of $N_C$ are impossible, although the behaviors of the $\sigma$ poles for the mild values around $N_C=3$ are found to be robust from different analyses.

\subsection{ The $N_C$ movements of light resonances from the dispersive Omn\`{e}s approach}

The dispersive Omn\`es function approach provides a stringent way to include the final-state strong interactions of the elastic two-body scattering~\cite{Omnes:1958hv}. In Refs.~\cite{Dai:2017uao,Dai:2018fmx}, the $N_C$ evolutions of the light-flavor meson resonances, including the $\sigma$, $f_0(980)$, $\rhov$ and $f_2(1270)$, and the semilocal duality in the finite energy sum rules are studied within the Omn\`es approach. To use the phases of the amplitudes, instead of the phase shifts, and to neglect the left-hand cuts, the single-channel $\pi\pi$ amplitudes up to 2~GeV are parameterized by the products of the polynomial terms and the Omn\`es functions, which will be simply denoted as Omn\`es amplitudes in later discussion. Through the matching between the Omn\`es amplitudes and the $\chi$PT ones in the low energy region, the parameters in the Omn\`es amplitudes can be given in terms of the $\chi$PT LECs. Since the $N_C$ scalings of the $\chi$PT LECs are known~\eqref{eq.nclecs}, it is possible to indirectly determine the $N_C$ scaling laws of the Omn\`es parameters. The $N_C$ movements of the various resonance poles can be then tracked~\cite{Dai:2017uao}. 
For the vector $\rhov$ and tensor $f_2(1270)$, they are confirmed to behave as the standard $\bar{q}q$ meson, with their masses approaching to constants and their widths decreasing as $1/N_C$. The $\sigma$-pole trajectory  is found to run deeply in the complex energy plane, with the mass roughly behaving as $O(1)$ and the width as $O(N_C)$ when increasing the values of $N_C$. It is argued that $\sigma$ could be a mixture between different sates, including molecule, tetraquark and $\bar{q}q$. The $f_0(980)$ pole, though running away from the real axis for $N_C<6$, finally tends to fall down to the real axis below the $K\bar{K}$ threshold for large values of $N_C$. It is advocated that the $f_0(980)$ can be a mixture of the $K\bar{K}$ molecule and the $s\bar{s}$ bare state~\cite{Dai:2017uao}. Following the method proposed in Ref.~\cite{RuizdeElvira:2010cs}, the semilocal duality is studied by varying the $N_C$ and is found to be satisfied up to large values of $N_C$. The effects from the $\rhov$ are found to be balanced by the $f_0(980)$, $f_0(1370)$ and $f_2(1270)$, while the $\sigma$ resonance plays little role. This result agree with the findings in Refs.~\cite{Guo:2012ym,Guo:2012yt}, where it is found that the $f_0(980)$, by turning into a bound state of 1~GeV at large $N_C$, is the one that compensates the vector  contributions. While in Ref.~\cite{RuizdeElvira:2010cs}, it is claimed that the $\sigma$, which eventually becomes a bound state around 1~GeV in the two-loop IAM, cancels the effect from the $\rhov$.

\subsection{Linear-sigma-model like study of the strange scalar resonances} 

The two light-flavor scalar resonances with a strange quark/antiquark, $\kappa$ and $K^*_0(1430)$, including their $N_C$ trajectories, are studied within the linear-sigma-model like framework in Ref.~\cite{Wolkanowski:2015jtc}. Both the derivative and non-derivative types of interaction operators are taken into account to calculate the decay widths and spectral functions, which in turn are used to obtain the phase shifts. The two resonances are generated by one bare state with mass around 1.1-1.3~GeV. It concludes that the $\kappa$ resonance could be a companion state of the $K^*_0(1430)$. To rescale the coupling constants in terms of $N_C$, the resonances poles of the $\kappa$ and $K^*_0(1430)$ with varying $N_C$ are then studied. The $K^*_0(1430)$ pole moves to the real axis for large $N_C$, behaving like the standard $\bar{q}q$ meson. In contrast, the pole corresponding to the $\kappa$ resonance moves deeply in the complex plane when increasing the $N_C$; hence, it provides another hint that $\kappa$ is unlikely dominated by the $\bar{q}q$ seed~\cite{Wolkanowski:2015jtc,Pelaez:2004xp}. 

\subsection{Light-flavor scalars from the unitarized-quark-model study }

Unitarized quark-model-like approach provides another type of theoretical frameworks to address the scalar resonances~\cite{Tornqvist:1982yv,Achasov:1994iu,vanBeveren:2006ua,Giacosa:2006tf,Zhou:2010ra,Lukashov:2019dir}, which usually needs to introduce a bare $q\bar{q}$ seed, in addition to the two-pNGB continuum contributions. The mass of the bare $q\bar{q}$ seed from these studies is found to be typically close to or even larger than 1~GeV.

In Ref~\cite{Zhou:2010ra}, the unitarized quark model simultaneously includes the $\bar{q}q$ bare states and the two-pNGB continuum contributions to study the light-flavor scalar meson spectra. The Adler zeros from the $\chi$PT are also taken into account as additional constraints. Large amount of experimental scattering data up to 2~GeV are fitted to fix the unknown parameters. A large number of scalar resonance poles are then obtained, including the isoscalars $\sigma$, $f_0(980)$, $f_0(1370)$, $f_0(1500)$, $f_0(1710)$, $f_0(2020)$, the isovectors $a_0(980)$,  $a_0(1450)$, $a_0(2020)$ and the strange scalars $\kappa$, $K^*_0(1430)$, $K^*_0(1950)$. It is found that when increasing the values of $N_C$ the resonance poles of $\sigma$, $f_0(980)$, $\kappa$ and $a_0(980)$ move away from the real axis, indicating that the bare  $\bar{q}q$ seeds play minor roles in these scalars. For all other heavier scalar resonances, they fall down exactly to the mass positions of the  $q\bar{q}$ seeds in the real axis at large values of $N_C$, implying that the $\bar{q}q$ seeds are crucial in their formations.

\subsection{ The $N_C$ evolutions of the axial-vector resonances}

The $N_C$-trajectory study is also extended to the axial-vector resonances in Ref.~\cite{Geng:2008ag}, where the axial-vector resonances $a_1(1260)$, $h_1(1170)$, $h_1(1380)$, $b_1(1235)$, $b_1(1285)$ and $K_1(1270)$ are simultaneously generated via the unitarization of the vector and pNGB scattering amplitudes. Only the contact interactions between the vectors and pNGBs are included in the study of Ref.~\cite{Geng:2008ag}, without considering any bare state exchange contribution. It turns out that all the poles of the dynamically generated axial-vector resonances move away from the real axis and run deeply in the complex energy plane, when increasing the values of $N_C$. Therefore it is concluded that all the axial-vector resonances, $a_1(1260)$, $h_1(1170)$, $h_1(1380)$, $b_1(1235)$, $b_1(1285)$ and $K_1(1270)$, studied in Ref.~\cite{Geng:2008ag}, are very unlikely to be the conventional $\bar{q}q$ mesons.

\subsection{$N_C$ dependences of the light resonances in elastic $\pi\pi$ scattering from the resonance chiral theory }

The elastic $\pi\pi$ scattering from the resonance chiral theory, which explicitly includes the bare resonance states and the pNGBs as the dynamical degrees of freedom, is explored to unravel the $N_C$ paths of the $\sigma$ and $\rhov$ resonances in Refs.~\cite{Sannino:1995ik,Nieves:2011gb}. The pioneer work of this type of studies to address the broad $\sigma$ resonance was carried out in Ref.~\cite{Sannino:1995ik}, which advocates a tetraquark interpretation of the bare $\sigma$ state, belonging to a subleading $1/N_C$ effect, with a bare mass around 500~MeV. In Ref.~\cite{Nieves:2011gb}, the main motivation to include the full propagators of the bare resonance exchanges is to keep all the leading $N_C$ terms without any truncation up to a specific chiral order. To impose the leading large $N_C$ high energy constraints in the chiral limit, the elastic $S$- and $P$-wave $\pi\pi$ scattering amplitudes are extrapolated to larger values of $N_C$, and the $N_C$ variations of the $\sigma$ and $\rhov$ poles are tracked in the complex energy plane. The $\rhov$ is robustly confirmed to become a $\bar{q}q$ at large $N_C$. Less clear conclusions are made for the $N_C$ movement of the $\sigma$ pole, since it is sensitive to the model parameters from different fit procedures~\cite{Nieves:2011gb}. This fact is also noticed in the chiral-limit IAM study in Ref.~\cite{Nieves:2009kh}, where the large cancellation of the LECs are found in the scalar isoscalar channel and a rather precise determination of the LECs would be needed to give definite conclusion for the $N_C$ fate of the $\sigma$. It is pointed out that the two-pion constituent dominates the $\sigma$ at $N_C=3$, but the $\bar{q}q$-like bare scalar state becomes important when $N_C\to\infty$~\cite{Nieves:2011gb}.

\subsection{$N_C$ spectra of light meson resonances in lattice QCD}

In the last decade there has been an important progress of lattice QCD  calculations on the $N_C$ dependence of the meson spectra~\cite{Lucini:2012gg,Lucini:2013qja}. However, most of the current lattice simulations on the $N_C$ study on the resonance spectra are still quite coarse. E.g. many of the lattice results on the spectra of the meson resonances with varying $N_C$ are obtained with the quenched  approximations, unphysically large quark  masses, or very small volumes~\cite{DelDebbio:2007wk,Hietanen:2009tu,Bali:2013kia,DeGrand:2016pur,Perez:2020fqn}. 

The quenched lattice QCD, which simplifies the numerical simulations by excluding the effects of the dynamical sea quarks, recovers the full QCD in the limit of $N_C\to\infty$, since the quark-loop contributions are $1/N_C$ suppressed. In the quenched case by neglecting the quark loops, the large $N_C$ QCD predicts that the subleading corrections for the hadronic observables behave as $1/N_C^2$, in contrast to the correction by including the dynamical fermions starting already at the order of $1/N_C$. By taking several finite values of $N_C$ in the simulations ranging from 3 to 17, the subleading corrections of the hadron masses from the quenched QCD studies nicely exhibit the quadratic scaling in the $1/N_C$  expansion~\cite{DelDebbio:2007wk,Bali:2013kia}. Currently the $N_C$ study of the hadron spectra from the unquenched lattice QCD with dynamical fermions is still sparse, and the exploratory unquenched simulation in Ref.~\cite{DeGrand:2016pur} seems confirming the expected $1/N_C$ scaling in several hadronic observables, including the masses and decay constants. However, larger finite volumes are clearly needed to reach definite conclusions~\cite{DeGrand:2016pur}. Moreover, up to date, most of lattice calculations focus on the the masses when varying $N_C$, and the discussions on the $N_C$ scalings of the decay widths are still absent. Recently, relatively finer unquenched lattice simulations have been done for the mass and decay constant of the pion, and the non-leptonic kaon decay process~\cite{Hernandez:2019qed,Donini:2020qfu}. Future unquenched lattice simulations with relatively large volumes by varying the numbers of $N_C$ could provide important constraints for the $\chi$PT community to further pin down the $N_C$ fates of the various resonances, specially the scalar ones.

\subsection{$N_C$ trajectories of the scalar charmed meson states}

The $N_C$ trajectories of the resonance poles are now widely focused on the light-flavor hadrons, and the discussions on the heavy-flavor resonances are still rare. In Ref.~\cite{Guo:2015dha}, the scalar charmed mesons are studied in the pNGB-$D$ scattering processes with different isospin and strangeness quantum numbers. Although the main topics of this note focus on the light-flavor hadrons, we briefly discuss the findings of the $N_C$ behaviors of the charmed scalar meson resonances in the former reference. The scalar charmed meson states, including $D^*_{s0}(2317)$ and $D^*_0(2400)$, are important objects in the hadron phenomenological study. The $N_C$ trajectories of their poles can provide valuable information about the internal structures for these two mesons. The algebraic approximation of the $N/D$ method is used to unitarize the NLO chiral amplitudes. The LECs and the subtraction constants are fitted to the lattice scattering lengths~\cite{Liu:2012zya} obtained at different quark masses. The physical $D^*_{s0}(2317)$ corresponds to a bound state pole below the $DK$ threshold. Two poles, with $I=1/2$ and strangeness zero,    corresponding to the $D^*_0(2400)$, are found in the complex energy plane. The lighter pole is located around 2.1~GeV and the heavier one is around 2.4~GeV. Interesting $N_C$ trajectory for the $D^*_{s0}(2317)$ is unveiled in Ref.~\cite{Guo:2015dha}. At $N_C=3$, the $D^*_{s0}(2317)$ is just a bound state pole at 2.3~GeV. When increasing the values of $N_C$ up to 6, this bound-state pole approaches to the $DK$ threshold, which becomes a virtual pole in the second RS for $N_C=7$. In the meantime, another virtual pole also appears near the threshold energy region, and the two virtual poles collide each other to become a pair of resonance poles when keeping increasing the values of $N_C$. The resonance poles seem going deep in the complex energy plane for large $N_C$. For the two poles corresponding to the $D^*_0(2400)$, both of them move away from the real axis in the complex energy plane when increasing the values of $N_C$~\cite{Guo:2015dha}. The $N_C$ behaviors of the $D^*_{s0}(2317)$ and $D^*_0(2400)$ poles imply that the $\bar{q}q$ constituents are unlikely to play dominant roles in their formation.

\subsection{Resonance observables with enhanced $1/N_C$ suppression}

In Refs.~\cite{Nieves:2009kh,Nebreda:2011cp}, the authors propose to directly study the resonance observables with enhanced $1/N_C$ suppression, namely the subleading $1/N_C$ correction includes at most the $1/N_C^2$ term compared to the leading one. It is nicely demonstrated in Ref.~\cite{Nieves:2009ez} that by performing the expansions of the real and imaginary parts of the inverse scattering $T$ matrix separately, it leads to the $1/N_C^{2n+1}$ types of power corrections in each expansion series. Equivalently the phase shift $\delta(M_R^2)$ and its derivative $\delta'(M_R^2)$, with the resonance pole $s_R=M_R^2-i M_R \Gamma_R$, are also found to fall in the aforementioned category of  observables~\cite{Nieves:2009ez}. Other observables with even further enhanced $1/N_C$ suppression are studied and scrutinized discussions using precise $\pi\pi$ and $K\pi$ phase shifts from the Roy dispersive analyses are carried out for the $\sigma$ and $\kappa$ resonances~\cite{Nebreda:2011cp}. The key finding is that the coefficients of the enhanced $1/N_C$ correction terms in the expansion series of the $\rho$ and $\kv$ channels turn out to be natural values around one or less, which are the expected results for the $q\bar{q}$ mesons. However the corresponding parameters in the $\sigma$ and $\kappa$ cases are found to be huge, around two orders larger than the ones from the vector channels, indicating that it is very unnatural to interpret the $\sigma$ and $\kappa$ as $q\bar{q}$ states~\cite{Nebreda:2011cp}. It should be stressed that in this kind of studies one does not really need to tune $N_C$. The $1/N_C$ simply offers an expansion parameter for the physical observables. This kind of approach does not rely on the unitarized chiral amplitudes and hence provides a different theoretical framework to gain insights into internal structures of the resonances based on the $1/N_C$ expansion.

\section{$N_C$ evolvements of the light meson dynamics in the $U(3)$ chiral theory}
\label{sec:u3}

If the number of colors is taken large, the quark loop graph which gives rise to the $U_A(1)$ anomaly is suppressed~\cite{ua1anomaly,Kaiser:2000gs}. Local non-invariance of the quark condensates under the $U(1)$ axial transformations leads to a ninth Goldstone boson in the large $N_C$ limit~\cite{ua1nc}.~\footnote{Witten argued that the instanton contribution~\cite{tHooft:1976rip} to the $\eta'$ mass scales as $e^{-N_C}$~\cite{Witten:1978bc} while the confinement contributions would give $m_{\eta'}^2\sim 1/N_C$. However, it is argued in Ref.~\cite{Schafer:2002af} that the contribution from larger instantons to the $\eta'$ can scale as $\sim 1/N_C$ as well.} The large $N_C$ treatment of the QCD $U_A(1)$ anomaly is specially attractive to the $\chi$PT practitioners, since the massive singlet $\eta_0$ state, whose large mass is believed to originate from the $U_A(1)$ anomaly, can be then systematically incorporated in the chiral effective theory. Although the mass squared $M_0^2$ of the singlet $\eta_0$ is large around 1~GeV$^2$ at $N_C=3$, it can be treated as a small expansion parameter for large $N_C$, since $M_0^2$ scales as $1/N_C$~\cite{ua1nc}. As a result the singlet $\eta_0$ would become massless and turn to the ninth Goldstone boson in the chiral and large $N_C$ limits, implying that the complete dynamical meson fields of the low energy QCD for $N_C\to\infty$ should include simultaneously the nine pNGBs $\pi, K, \eta$ and $\eta'$, not just the  octet states. It should be pointed out that all the previous studies on the $N_C$ movements of the light meson resonances in the former section have neglected the singlet $\eta_0$ as an explicit degree of freedom. The $U(3)$ chiral theory explicitly includes the nine pNGBs $\pi, K, \eta$ and $\eta'$~\cite{ua1nc}, dynamically taking the QCD $U_A(1)$ anomaly into account, therefore it provides a more appropriate theoretical framework to investigate the $N_C$ evolution of the various resonances. In Refs.~\cite{Guo:2011pa,Guo:2012ym,Guo:2012yt}, systematical calculations up to the one-loop level have been carried out in the framework of the $U(3)$ chiral theory with explicit resonance exchanges, which are further used in many other phenomenological discussions after including the finite-volume and the finite-temperature effects~\cite{Guo:2016zep,Gao:2019idb}. The discussions of the $N_C$ trajectories of the light-flavor meson resonances and related physical quantities are thoroughly addressed in detail. We briefly recapitulate the key results here.

To set up the consistent power counting rules to simultaneously include the nine pNGBs $\pi, K, \eta$ and $\eta'$, the $U(3)$ $\chi$PT introduces the triple-expansion formalism, also simply denoted as $\delta$ expansion, namely $\delta\sim p^2 \sim m_q \sim 1/N_C$~\cite{HerreraSiklody:1996pm,Kaiser:2000gs}. The LO $U(3)$ chiral Lagrangian consists of three independent $O(\delta^{0})$ operators 
\begin{eqnarray}\label{eq.lag1}
 \mL_{0}=  \frac{F^2}{4}\bra u_\mu u^\mu \ket+
\frac{F^2}{4}\bra \chi_+ \ket + \frac{F^2}{3} M_0^2 \ln^2{\det u} \,,
\end{eqnarray}
where the chiral tensors are given by 
\begin{eqnarray}\label{eq.efbb}
&& U =  u^2 = e^{i\frac{ \sqrt2\Phi}{ F}}\,, \qquad \chi = 2 B (s + i p) \,,\qquad \chi_\pm  = u^\dagger  \chi u^\dagger  \pm  u \chi^\dagger  u \,, \nn\\
&& u_\mu = i u^\dagger  D_\mu U u^\dagger \,, \qquad  D_\mu U \, =\, \partial_\mu U - i (v_\mu + a_\mu) U\, + i U  (v_\mu - a_\mu) \,,
\end{eqnarray}
and the matrix of the nonet pNGB fields reads
\begin{equation}\label{eq.phi9}
\Phi \,=\, \left( \begin{array}{ccc}
\frac{1}{\sqrt{2}} \pi^0+\frac{1}{\sqrt{6}}\eta_8+\frac{1}{\sqrt{3}} \eta_0 & \pi^+ & K^+ \\ \pi^- &
\frac{-1}{\sqrt{2}} \pi^0+\frac{1}{\sqrt{6}}\eta_8+\frac{1}{\sqrt{3}} \eta_0   & K^0 \\  K^- & \overline{K}^0 &
\frac{-2}{\sqrt{6}}\eta_8+\frac{1}{\sqrt{3}} \eta_0
\end{array} \right)\,.
\end{equation}
$F$ is the LO pNGB decay constant, with the normalization $F_\pi=92.1$~MeV. The QCD $U_A(1)$ anomaly is given by the last term in Eq.~\eqref{eq.lag1}, which provides the LO mass $M_0$ to the singlet $\eta_0$ even at the chiral limit. The leading $N_C$ scaling for $M_0^2$ is $O(1/N_C)$~\cite{ua1nc}. Within the $\delta$-expansion framework, the order of $M_0^2$ is counted the same as the $m_\pi^2, m_K^2$ and $m_{\eta_8}^2$. At leading order, the mixing between the physical $\eta$, $\eta'$ and the $\eta_0$, $\eta_8$ can be described by one-mixing-angle formula 
\begin{eqnarray}\label{eq.lomixing}
\left(\begin{array}{c} \eta \\ \eta' \end{array}\right) \quad=\quad
\left(\begin{array}{cc} \cos\theta & - \sin\theta \\  \sin\theta & \cos\theta \end{array}\right) \quad
\left(\begin{array}{c} \eta_8 \\ \eta_0\end{array}\right) \,.
\end{eqnarray}
Beyond the leading order, one usually needs to introduce the two-mixing-angle scheme and interested readers are referred to Refs.~\cite{Guo:2015daa,Gu:2018swy} for a complete NNLO $U(3)$ $\chi$PT study of the $\eta$-$\eta'$ mixing. By diagonalizing the $\eta_0-\eta_8$ mixing term at leading order of Eq.~\eqref{eq.lag1}, the LO masses for the physical $\eta$ and $\eta'$ states and their mixing angle $\theta$ take the form~\cite{Guo:2011pa}
 \begin{eqnarray}
m_{\eta}^2 &=& \frac{M_0^2}{2} + m_K^2
- \frac{\sqrt{M_0^4 - \frac{4 M_0^2 \Delta^2}{3}+ 4 \Delta^4 }}{2} \,, \label{eq.defmetab2}  \\
m_{\eta'}^2 &=& \frac{M_0^2}{2} + m_K^2
+ \frac{\sqrt{M_0^4 - \frac{4 M_0^2 \Delta^2}{3}+ 4 \Delta^4 }}{2} \,, \label{eq.defmetaPb2}  \\
\sin{\theta} &=& -\left( \sqrt{1 +
\frac{ \big(3M_0^2 - 2\Delta^2 +\sqrt{9M_0^4-12 M_0^2 \Delta^2 +36 \Delta^4 } \big)^2}{32 \Delta^4} } ~\right )^{-1}\,,
\label{eq.deftheta0}
\end{eqnarray}
where all the masses here correspond to their LO values and $\Delta^2 = m_K^2 - m_\pi^2$.

\begin{figure}[htbp]
\begin{center}
\includegraphics[angle=-90,width=0.98\columnwidth]{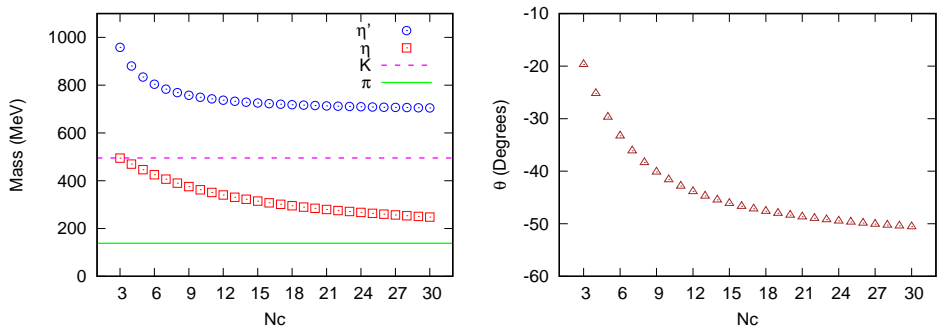}  
\caption{ The LO results for the $N_C$ evolutions of the pNGBs' masses (left panel) and the $\eta$-$\eta'$ mixing angle (right panel).} \label{fig.ncpsmassandtheta}
\end{center}
\end{figure}

Already at LO, it is interesting to illustrate the $N_C$ evolutions of the masses of the pNGBs and the $\eta$-$\eta'$ mixing angle. To take the physical values for the masses of the pion and kaon and fix $M_0=820.0$~MeV at $N_C=3$ from the recent determination~\cite{Gu:2018swy}, the $N_C$ variations of the pNGBs' masses and the mixing angle are shown in Fig.~\ref{fig.ncpsmassandtheta} for a wide range of $N_C$ from 3 to 30. The most obvious changes happen for the $m_\eta$ and the LO mixing angle $\theta$, which significantly reduces to around 250~MeV and approaches to the ideal mixing value around $-50^{\circ}$ respectively, when increasing the values of $N_C$ up to 30. Due to the large mass of the kaon, the mass of the $\eta'$ still remains quite large around 700~MeV for $N_C=30$. When including the higher order contributions~\cite{Guo:2011pa,Guo:2012ym,Guo:2012yt,Guo:2015xva}, the $N_C$ evolutions of the masses and the LO mixing angle in Fig.~\ref{fig.ncpsmassandtheta} are found to be barely affected, e.g. see Figs.~8 and 9 of Ref.~\cite{Guo:2011pa}. In contrast, the LO masses of the $\pi, K$ and $\eta$ will be simply constants within the $SU(3)$ $\chi$PT. To our knowledge,  Refs.~\cite{Guo:2011pa,Guo:2012ym,Guo:2012yt} are the first works in literature to take into account the striking $N_C$ evolutions of the pNGB masses and the LO mixing angle $\theta$ to determine the $N_C$ trajectories of the resonance poles.

The two-pNGB scattering amplitudes have been fully calculated up to the one-loop level~\cite{Guo:2011pa,Guo:2012ym,Guo:2012yt}, whose partial-wave amplitudes are then unitarized with an approximated $N/D$ approach. In the on-shell approximation of the $N/D$ method~\cite{Oller:1998zr,Oller:2000fj,Oller:1999me,Lacour:2009ej}, the left-hand contributions from the crossed channels are perturbatively included by matching the unitarized partial-wave amplitudes with their perturbative expressions calculated at one loop in $U(3)$ $\chi$PT~\cite{Guo:2011pa,Guo:2012ym,Guo:2012yt}. The $N/D$ approach overcomes the IAM problem when the inverse of the matrix $T_2-T_4$ in Eq.~\eqref{eq.iam} is singular. The explicit bare resonance contributions within the resonance chiral theory~\cite{Ecker:1988te}, including the scalar octet plus singlet, the vector nonet and the pseudoscalar resonance nonet, are considered both in the $s$- and crossed channels in the calculation~\cite{Guo:2011pa,Guo:2012ym,Guo:2012yt}. Vast amount of experimental scattering data, including the phase shifts and inelasticities of the $\pi\pi$ and $K\pi$ scattering with various isospin and angular momenta numbers, and also the $\pi\eta$ event distributions, are fitted to obtain the unknown parameters in our theoretical models, such as the subtraction constants arising from the unitarization procedure, the bare resonance masses and couplings, and two pure $U(3)$ LECs that can not be contributed by the considered resonance Lagrangians. By extrapolating the unitarized partial-wave amplitudes into the complex energy plane, the various resonance poles are searched in different Riemann sheets. We have found the poles for a large number of light-flavor resonances below 1.5~GeV, including the isoscalar scalar resonances $\sigma$, $f_0(980)$ and $f_0(1370)$, the isovector scalar resonances $a_0(980)$ and $a_0(1450)$, the strange scalar resonances $\kappa$ and $K_0^*(1430)$, and the vector resonances $\rhov$, $K^*(892)$ and $\phi(1020)$. Their masses and widths~\cite{Guo:2012yt} are nicely compatible with the world average values from PDG~\cite{Tanabashi:2018oca}.

Following the similar recipes in Eqs.~\eqref{eq.linc1}, \eqref{eq.linc0} or \eqref{eq.lincfpi}, the leading $N_C$ scaling of the bare resonance parameters including the masses and couplings is implemented in Ref.~\cite{Guo:2011pa}, where we only consider the values of the bare parameters determined from the fits during the $N_C$ scaling procedure. Later on Ref.~\cite{Nieves:2011gb} advocated imposing the leading $N_C$ scaling of the bare resonance parameters by also additionally taking into account the high energy constraints and the relations dictated by the large $N_C$ QCD. In Refs.~\cite{Guo:2012ym,Guo:2012yt}, we have explicitly taken into account such constraints on some important couplings and masses when varying $N_C$. Regarding the subtraction constants introduced in the unitarization procedure, it is argued in Ref.~\cite{Guo:2011pa} that they should behave as constants at the leading $N_C$ order, because any variation in them due to a change of the subtraction point in the unitarity loop functions is $O(N_C^0)$. In addition, the $N_C$ variations of the pNGBs' masses and the mixing angle, e.g. those in Fig.~\ref{fig.ncpsmassandtheta}, which are ignored in the previous $N_C$ studies, are carefully considered in the determination of the resonance pole trajectories for a wide range of $N_C$ from 3 to 30.

To qualitatively clarify the effects of the characteristic $N_C$ variations of the $U(3)$ chiral theory, as opposed to the $SU(3)$ case, we have proposed a procedure to mimic the $SU(3)$ $\chi$PT to study the $N_C$ evolutions of the various resonance pole positions by using the $U(3)$ amplitudes. The procedure includes three parts: (a) to fix the LO mixing angle $\theta=0$ throughout; (b) to freeze the $m_\pi, m_K, m_\eta$ at their physical values; (c) to fix the mass of the  singlet $\eta_0$ by its LO result in Eq.~\eqref{eq.defmetaPb2}. The $N_C$ trajectories from this procedure will be denoted as $SU(3)$ in the following discussions.

\begin{figure}[htbp]
\begin{center}
\includegraphics[angle=-90,width=0.98\columnwidth]{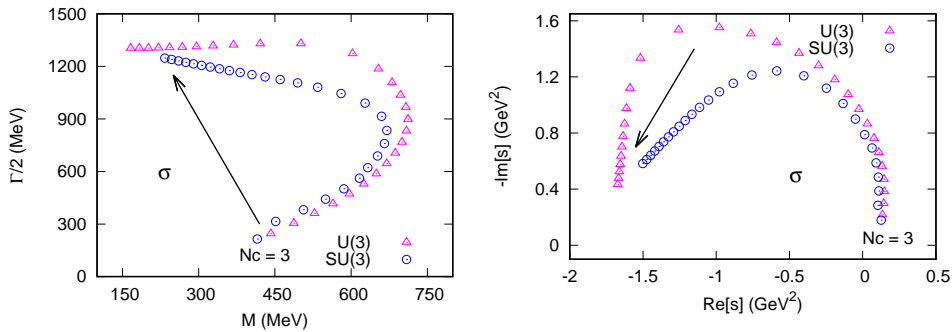}  
\caption{ The movements of the $\sigma$ poles with varying $N_C$ from 3 to 30~\cite{Guo:2012yt}. See the text for details about the differences of the $U(3)$ and $SU(3)$ results. The left panel shows the pole positions in the energy plane and the right panel gives the pole movements in the energy squared $s$ plane. } \label{fig.ncsigma}
\end{center}
\end{figure}

\begin{figure}[htbp]
\begin{center}
\includegraphics[angle=-90,width=1.0\columnwidth]{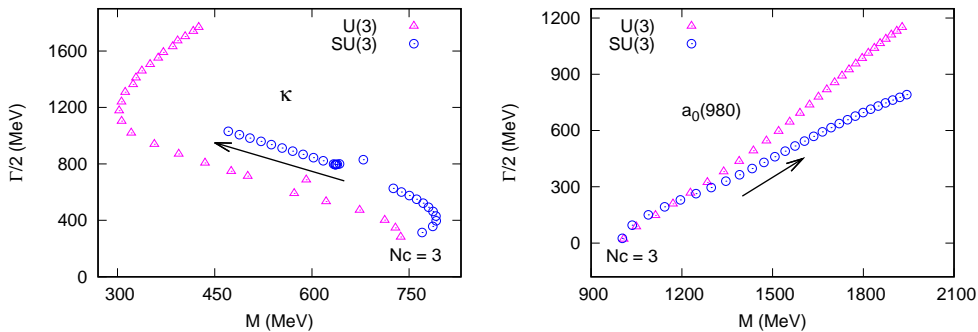}  
\caption{ The movements of the $\kappa$ and $a_0(980)$ poles in the complex energy plane with varying $N_C$ from 3 to 30~\cite{Guo:2012yt}. See the text for details about the differences of the $U(3)$ and $SU(3)$ results.} \label{fig.nca0kappa}
\end{center}
\end{figure}

The $N_C$ paths of the $\sigma$ resonance are illustrated in Fig.~\ref{fig.ncsigma}~\cite{Guo:2012yt}. First, the physical $\sigma$ pole at $N_C=3$ is close to those from the use of Roy-like equations~\cite{Tanabashi:2018oca}. To identify the real and imaginary parts of the pole position as the mass and half width, respectively, the movements of the $\sigma$ pole with varying $N_C$ are shown in the left panel of Fig.~\ref{fig.ncsigma}. The pole trajectories in the complex $s$ plane are shown in the right panel. When increasing the values of $N_C$ up to around 15, the $\sigma$ pole position keeps running away from the real axis in the energy squared $s$ plane. For $N_C>15$, its pole starts to move towards the real axis, but in our present study the $\sigma$ pole tends to move to the negative real $s$ axis, instead of the positive one, implying a loss of relevance of the physical signal attached to the $\sigma$ resonance at large $N_C$ within our approach. This also implies that the dominant components in the $\sigma$ resonance are not the $\bar{q}q$. This result is roughly similar with the one-loop $SU(3)$ IAM study~\cite{Pelaez:2003dy,Pelaez:2004xp}. The $\sigma$ poles from the $U(3)$ and $SU(3)$ cases are quantitatively close for $N_C\le 10$ and for large values of $N_C$ the two schemes still give qualitatively similar curves.

Regarding the pole variations of the $\kappa$ and $a_0(980)$~\cite{Guo:2012yt}, as shown in Fig.~\ref{fig.nca0kappa}, roughly speaking they share comparable trends as the $\sigma$, meaning that their poles tend to move deeper into the complex energy plane when increasing the values of $N_C$, instead of falling down to the real axis. According the $N_C$ behaviors in Fig.~\ref{fig.nca0kappa}, it is unlikely that there are important $\bar{q}q$ components inside the $\kappa$ and $a_0(980)$ at large $N_C$. This seems also consistent with the initial setups of our theoretical models, since the $\sigma$, $a_0(980)$ and $\kappa$ do not have any corresponding bare state in the resonance chiral Lagrangians and they are mainly generated from the strong interactions among the pNGBs. Both the $U(3)$ and $SU(3)$ treatments give qualitatively comparative $N_C$ trajectories to the $\kappa$ and $a_0(980)$, though the deviations for the $\kappa$ poles are relatively larger. It is pointed out that the $N_C$ curves for the $\kappa$ and $a_0(980)$ are also somewhat sensitive to the different fit strategies. For example, one can see some obvious differences of the $\kappa$ trajectories between the results in Refs.~\cite{Guo:2012yt,Guo:2011pa}, although in both cases the $\kappa$ pole dilutes deeply into the  complex energy plane.

\begin{figure}[htbp]
\begin{center}
\includegraphics[angle=-90,width=0.98\columnwidth]{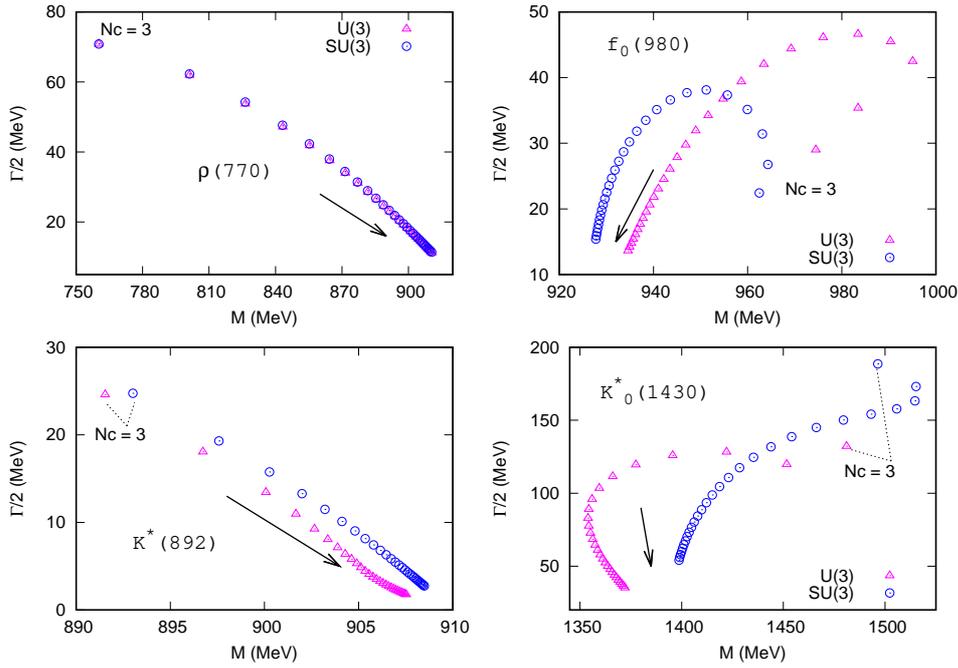}  
\caption{ The movements of the resonance poles for the $\rhov$, $f_0(980)$, $\kv$ and $K^*_0(1430)$ in the complex energy plane with varying $N_C$ from 3 to 30~\cite{Guo:2012yt}. See the text for details about the differences of the $U(3)$ and $SU(3)$ results.} \label{fig.nckvksrhof0}
\end{center}
\end{figure}

The $N_C$ trajectories of the resonances $\rhov$ and $f_0(980)$ in the coupled $\pi\pi$ scattering, and the $\kv$ and $K^*_0(1430) $ in the coupled $K\pi$ scattering, are illustrated in Figs.~\ref{fig.nckvksrhof0}~\cite{Guo:2012yt}. A striking difference of the $N_C$ pole movements between the $\sigma,\kappa,a_0(980)$ in Figs.~\ref{fig.ncsigma}~\ref{fig.nca0kappa} and the $\rhov,f_0(980),\kv,K^*_0(1430)$ in Fig.~\ref{fig.nckvksrhof0} is that the latter ones approach to the positive real axis in the complex energy plane for large values of $N_C$, which is a characteristic feature of a standard $\bar{q}q$ meson. The vector resonances $\rhov$ and $\kv$, exhibit perfect behaviors for $\bar{q}q$-like mesons, with their masses approaching to a constant and their widths vanishing precisely as $1/N_C$ when increasing the values of $N_C$. The differences caused by the $U(3)$ and $SU(3)$ theories are small for the vector resonances, in particular the deviations for the $\rhov$ are almost invisible for all values of $N_C$. The  $f_0(980)$ in our study receives important contributions both from the $\bar{K}K$ channel and a bare singlet scalar state, whose mass turns out to be around 1~GeV from the fits~\cite{Guo:2011pa,Guo:2012yt}. Were the role played by the bare singlet scalar dominant, one would expect that the $N_C$ paths of the $f_0(980)$ look similar to the vector ones. In the actual situation, the widths of the $f_0(980)$ tend to increase for $N_C<8$, which hints that the $\bar{q}q$ component is subdominant and probably the molecular component of the $\bar{K}K$ is the dominant one in this region. For the case of $N_C\ge 8$ the $f_0(980)$ pole moves towards the real axis with decreasing width and more or less constant mass, indicating that the $\bar{q}q$ type bare state starts to play the predominant role at large $N_C$. For the heavier excited scalar resonances $K^*_0(1430)$, $f_0(1370)$ and $a_0(1450)$, they receive important contributions from the bare scalar octet in the resonance chiral Lagrangian, whose mass is determined to be around 1.4~GeV from the fits~\cite{Guo:2011pa,Guo:2012yt}. For large values of $N_C$, their pole positions all run to the real axis with vanishing widths and we explicitly show the results for the $K^*_0(1430)$ in Fig.~\ref{fig.nckvksrhof0}. Other $N_C$ curves for the $f_0(1370)$ and $a_0(1450)$ are quite similar and can be found in Refs.~\cite{Guo:2011pa,Guo:2012yt}. A general feature for the heavier excited scalar resonances is that their $N_C$ trajectories are moderately sensitive to the different treatments of the $U(3)$ and $SU(3)$ theories. Although different prescriptions of the $U(3)$ and $SU(3)$ will not alter the final trends of the pole movements, they do lead to obviously different $N_C$ paths for the heavier excited scalar resonances, as one can see the panels of the $f_0(980)$ and $K^*_0(1430)$. Similar conclusions are also obtained for the $f_0(1370)$ and $a_0(1450)$~\cite{Guo:2011pa,Guo:2012yt}. 

The interesting $N_C$ evolutions of the various resonance poles will definitely influence other related physical quantities. We have focused on the semilocal duality in the $\pi\pi$ scattering and the Weinberg spectral sum rules in the scalar and pseudoscalar sectors in Refs.~\cite{Guo:2012yt}. Their variations with different values of $N_C$ are specially paid attention to. For the fixed-$t$ meson-meson scattering amplitudes, the semilocal duality refers to the duality relation between the Regge theory and the hadronic system. In   $\pi\pi$ scattering, the semilocal duality relation provides a useful object to study the interplays between the vector and scalar resonances, and it can be explicitly verified through the finite-energy sum rules~\cite{RuizdeElvira:2010cs}. Due to the absence of the $s$-channel resonances in the $\pi\pi$ amplitude with $I=2$, the Regge theory predicts vanishing results for the isotensor amplitude. The duality relation then requires a strong cancellation of the resonances from the crossed channels in the hadronic side. In Refs.~\cite{Guo:2012ym,Guo:2012yt}, we have used the semilocal duality to study the possible relations between the scalar ($\sigma,f_0(980),f_0(1370)$) and the vector ($\rhov$) resonances. At $N_C=3$, to fulfill the duality relation the $\sigma$ resonance generally plays important roles in balancing the $\rhov$. For large values of $N_C$, it is found that a scalar resonance at around 1~GeV is needed to balance the $\rhov$ contribution to satisfy the semilocal duality relation~\cite{RuizdeElvira:2010cs}. In our case, the contributions from the $\sigma$ gradually fade away, and it is the $f_0(980)$ that finally becomes the one to cancel the effect from the $\rhov$~\cite{Guo:2012yt}, through the contribution to its composition of an elementary state introduced as a bared field, as  discussed above. 

On the other hand, the Weinberg spectral sum rules of the $SS-PP$ types, with $S$ the scalar currents and $P$ the pseudoscalar currents, offer another theoretical tool to examine the relations between the scalar and pseudoscalar resonances. The scalar spectral functions with different flavor contents are calculated via the unitarized scalar form factors of the two-pNGB states~\cite{Guo:2012ym,Guo:2012yt,Meissner:2000bc,Oller:2000ug}. Similarly the pseudoscalar spectral functions are constructed with the pseudoscalar form factors contributed by the single-state exchanges of the pNGBs and the pseudoscalar resonances~\cite{Albaladejo:2010tj}. According to the results of the perturbative QCD and OPE, the $SS-PP$ spectral function integrals in the high energy region tend to vanish in the chiral limit. This in turn implies the cancellation between the scalar and pseudoscalar spectral integrals in the nonperturbative region, if the $SS-PP$ Weinberg sum rules hold. By properly performing the chiral extrapolation, the spectral sum rules are simultaneously examined in the physical quark-mass situation and in the chiral-limit case. Three different flavor types of the spectral functions are considered, namely the $SU(3)$ singlet current $(\bar{u}\Gamma u+\bar{d}\Gamma d+\bar{s}\Gamma s)/\sqrt{3}$, the $SU(3)$ octet $(\bar{u}\Gamma u+\bar{d}\Gamma d-2\bar{s}\Gamma s)/\sqrt{6}$ and the isovector $(\bar{u}\Gamma u-\bar{d}\Gamma d)/\sqrt{2}$, being $\Gamma=1$ for the scalar condensates and $\Gamma=\gamma_5$ for the pseudoscalar condensates. The updated fits in Ref.~\cite{Guo:2012yt} indicate that the $SS-PP$ Weinberg spectral sum rules are well satisfied for a wide range values of $N_C$, with the violation at the level less than  16\%~\cite{Guo:2012yt}. It is interesting to point out that the mechanisms for the fulfillment of the $SS-PP$ Weinberg sum rules at different stages in the evolution with $N_C$ are different. For the $SU(3)$ singlet spectral sum rule, both the $\sigma$ and $f_0(980)$ scalar resonances are found to be important to compensate the contributions from the pseudoscalar mesons at $N_C=3$. For the $SU(3)$ octet case, in addition to the $\sigma$ and $f_0(980)$, the $f_0(1370)$ also gives important contributions. For the isovector $(\bar{u}\Gamma u-\bar{d}\Gamma d)/\sqrt{2}$ case at $N_C=3$, it turns out that the rather prominent $a_0(980)$ peak gives more important contributions than the $a_0(1450)$ to the scalar spectral integral. However, as shown in Figs.~\ref{fig.ncsigma}~\ref{fig.nca0kappa}~\ref{fig.nckvksrhof0}, different scalar resonances show clearly different $N_C$ trajectories, implying that their contributions to the scalar spectral functions will also vary with different values of $N_C$. More precisely speaking, the effects of the $\sigma$ and $a_0(980)$, though important at $N_C=3$, start to diminish for large values of $N_C$, while the contributions from the $f_0(980)$, $f_0(1370)$ and $a_0(1450)$ become crucial in the large $N_C$ limit. Once the high-energy constraints on the resonance bare couplings are properly imposed, the $SS-PP$ Weinberg spectral sum rules are found to be perfectly fulfilled in the chiral limit at large $N_C$~\cite{Guo:2012ym,Guo:2012yt}.

\section{Summary and conclusions}

In this work, we give an extensive review on the variations of the pole positions of various light-flavor meson resonances and related physical quantities by varying the number of colors $N_C$ of QCD. Although it is difficult to give quantitative conclusions on the internal structures, the $N_C$ trajectories of the resonance pole positions, namely the movements of their masses and widths with varying $N_C$, can provide intuitive and, some times, even compelling results to discern the interior constituents of resonances. 

Many theoretical studies from different research groups, including the inverse amplitude method, the $N/D$ approach, unitarized quark models, dispersive Omn\`es function, linear-sigma-model like scheme, etc., have been proposed to examine the $N_C$ trajectories of  meson resonance poles. A general agreement from different models on the $N_C$ behaviors of the vector resonances, such as the $\rhov$ and $\kv$, has been reached. As expected for a conventional $\bar{q}q$ meson, the masses approach to constants and the widths nicely decrease as $1/N_C$ when taking larger values of $N_C$. The movement in $N_C$ of the vector-resonance poles can be considered as a calibration for the model. 

The situation for the scalar resonances is more subtle. It is enlightening to elaborate on their behaviors in different $N_C$ regions, namely the ones near $N_C=3$ and those in the large $N_C$ case. A more or less robust behavior revealed by different groups for the broad $\sigma$ or $\kappa$ is that their poles seem not falling down to the real axis for $N_C$ not far away from its physical value. However, the trajectories of the $\sigma$ and $\kappa$ resonances at large values of $N_C$ turn out to be sensitive to many factors, such as different fit strategies, higher order low energy constants, different unitarization models, etc. It is found in some works that the $\sigma$ and $\kappa$ poles fade away at large $N_C$. In other situations, such as the two-loop inverse-amplitude-method case, the $\sigma$ pole finally turns back to the positive real axis at the large $N_C$, indicating that there is a $q\bar{q}$ seed in the $\sigma$ with a mass around 1~GeV,  which is rather different from the physical mass of the $\sigma$. Compared to the sharp  $N_C$ trajectories for the vector resonances, all of these features found for the broad $\sigma$ and $\kappa$ imply that they do not seem to correspond to the standard $q\bar{q}$ states, so that the $q\bar{q}$ seeds are unlikely the dominant constituents.

The presence of the $\bar{K}K$ threshold makes the studies of the $f_0(980)$ and $a_0(980)$ more complicated, and the $N_C$ movements of the two resonance poles from different studies also show different trends. The recent $U(3)$ study on the heavier excited scalar resonances $f_0(1370)$, $K^*_0(1430)$ and $a_0(1450)$ show that their pole positions move towards the real axis with vanishing widths for large $N_C$. The finite-volume energy levels related to the scalar resonances $\sigma$ and $f_0(980)$ in the unquenched lattice QCD nowadays have been obtained with great precision with $N_C=3$~\cite{Briceno:2016mjc,Briceno:2017qmb}. Future similar unquenched lattice simulations but with different values of $N_C$ could provide crucial hints of the internal structures of the scalar resonances.

\section*{Acknowledgments}
I would like to thank J.~A.~Oller and J.~Ruiz de Elvira for the careful reading of the manuscript and useful suggestions. This work is partially funded by the Natural Science Foundation of China under Grant Nos.~11975090 and ~11575052, the Natural Science Foundation of Hebei Province under Contract No.~A2015205205, and the Fundamental Research Funds for the Central Universities.


\begin{thebibliography}{}

\bibitem{Pelaez:2015qba}
J.~R.~Pelaez,
Phys. Rept. \textbf{658}, 1 (2016)
doi:10.1016/j.physrep.2016.09.001
[arXiv:1510.00653 [hep-ph]].

\bibitem{Yao:2020bxx}
D.~L.~Yao, L.~Y.~Dai, H.~Q.~Zheng and Z.~Y.~Zhou,
[arXiv:2009.13495 [hep-ph]].

\bibitem{Oller:2019opk}
J.~A.~Oller,
Prog. Part. Nucl. Phys. \textbf{110}, 103728 (2020)
doi:10.1016/j.ppnp.2019.103728
[arXiv:1909.00370 [hep-ph]].

\bibitem{Oller:2020guq}
J.~A.~Oller,
Symmetry \textbf{12}, no.7, 1114 (2020)
doi:10.3390/sym12071114
[arXiv:2005.14417 [hep-ph]].

\bibitem{Weinberg:1962hj} 
  S.~Weinberg,
  Phys.\ Rev.\  {\bf 130}, 776 (1963).
  doi:10.1103/PhysRev.130.776
  
  \bibitem{Baru:2003qq} 
  V.~Baru, J.~Haidenbauer, C.~Hanhart, Y.~Kalashnikova and A.~E.~Kudryavtsev,
  Phys.\ Lett.\ B {\bf 586}, 53 (2004)
  doi:10.1016/j.physletb.2004.01.088
  [hep-ph/0308129].
  
\bibitem{Hanhart:2011jz} 
  C.~Hanhart, Y.~S.~Kalashnikova and A.~V.~Nefediev,
  Eur.\ Phys.\ J.\ A {\bf 47}, 101 (2011)
  doi:10.1140/epja/i2011-11101-9
  [arXiv:1106.1185 [hep-ph]].
  
\bibitem{Hyodo:2011qc} 
  T.~Hyodo, D.~Jido and A.~Hosaka,
  Phys.\ Rev.\ C {\bf 85}, 015201 (2012)
  doi:10.1103/PhysRevC.85.015201
  [arXiv:1108.5524 [nucl-th]].
  
\bibitem{Aceti:2012dd} 
  F.~Aceti and E.~Oset,
  Phys.\ Rev.\ D {\bf 86}, 014012 (2012)
  doi:10.1103/PhysRevD.86.014012
  [arXiv:1202.4607 [hep-ph]].
  
\bibitem{Sekihara:2014kya} 
  T.~Sekihara, T.~Hyodo and D.~Jido,
  PTEP {\bf 2015}, 063D04 (2015)
  doi:10.1093/ptep/ptv081
  [arXiv:1411.2308 [hep-ph]].


\bibitem{Guo:2015daa} 
  Z.~H.~Guo and J.~A.~Oller,
  Phys.\ Rev.\ D {\bf 93}, no. 9, 096001 (2016)
  doi:10.1103/PhysRevD.93.096001
  [arXiv:1508.06400 [hep-ph]].
  
\bibitem{Oller:2017alp} 
  J.~A.~Oller,
  Annals Phys.\  {\bf 396}, 429 (2018)
  doi:10.1016/j.aop.2018.07.023
  [arXiv:1710.00991 [hep-ph]].

\bibitem{Matuschek:2020gqe}
I.~Matuschek, V.~Baru, F.~K.~Guo and C.~Hanhart,
Eur. Phys. J. A \textbf{57}, no.3, 101 (2021)
doi:10.1140/epja/s10050-021-00413-y
[arXiv:2007.05329 [hep-ph]].

\bibitem{Gao:2018jhk}
R.~Gao, Z.~H.~Guo, X.~W.~Kang and J.~A.~Oller,
Adv. High Energy Phys. \textbf{2019}, 4651908 (2019)
doi:10.1155/2019/4651908
[arXiv:1812.07323 [hep-ph]].

\bibitem{Hanhart:2008mx}
C.~Hanhart, J.~R.~Pelaez and G.~Rios,
Phys. Rev. Lett. \textbf{100}, 152001 (2008)
doi:10.1103/PhysRevLett.100.152001
[arXiv:0801.2871 [hep-ph]].

\bibitem{Hanhart:2014ssa}
C.~Hanhart, J.~R.~Pelaez and G.~Rios,
Phys. Lett. B \textbf{739}, 375-382 (2014)
doi:10.1016/j.physletb.2014.11.011
[arXiv:1407.7452 [hep-ph]].

\bibitem{Liu:2012zya}
L.~Liu, K.~Orginos, F.~K.~Guo, C.~Hanhart and U.~G.~Meissner,
Phys. Rev. D \textbf{87}, no.1, 014508 (2013)
doi:10.1103/PhysRevD.87.014508
[arXiv:1208.4535 [hep-lat]].

\bibitem{Torres:2014vna}
A.~Mart\'\i{}nez Torres, E.~Oset, S.~Prelovsek and A.~Ramos,
JHEP \textbf{05}, 153 (2015)
doi:10.1007/JHEP05(2015)153
[arXiv:1412.1706 [hep-lat]].

\bibitem{Guo:2015dha}
Z.~H.~Guo, U.~G.~Mei\ss{}ner and D.~L.~Yao,
Phys. Rev. D \textbf{92}, no.9, 094008 (2015)
doi:10.1103/PhysRevD.92.094008
[arXiv:1507.03123 [hep-ph]].


\bibitem{RuizdeElvira:2017aet}
J.~Ruiz de Elvira, U.~G.~Mei\ss{}ner, A.~Rusetsky and G.~Schierholz,
Eur. Phys. J. C \textbf{77}, no.10, 659 (2017)
doi:10.1140/epjc/s10052-017-5237-3
[arXiv:1706.09015 [hep-lat]].

\bibitem{Ren:2012aj}
X.~L.~Ren, L.~S.~Geng, J.~Martin Camalich, J.~Meng and H.~Toki,
JHEP \textbf{12}, 073 (2012)
doi:10.1007/JHEP12(2012)073
[arXiv:1209.3641 [nucl-th]].

\bibitem{Aoki:2019cca}
S.~Aoki \textit{et al.} [Flavour Lattice Averaging Group],
Eur. Phys. J. C \textbf{80}, no.2, 113 (2020)
doi:10.1140/epjc/s10052-019-7354-7
[arXiv:1902.08191 [hep-lat]].

\bibitem{largenc}
G.~'t Hooft,
Nucl. Phys. B \textbf{72}, 461 (1974)
doi:10.1016/0550-3213(74)90154-0; 

G.~'t Hooft,
Nucl. Phys. B \textbf{75}, 461-470 (1974)
doi:10.1016/0550-3213(74)90088-1; 


E.~Witten,
Nucl. Phys. B \textbf{160}, 57-115 (1979)
doi:10.1016/0550-3213(79)90232-3; 

\bibitem{Coleman:1985rnk}
S.~Coleman, {\it Aspects of Symmetry: Selected Erice Lectures}. Cambridge University Press (1985). 
doi:10.1017/CBO9780511565045


\bibitem{Weinberg:2013cfa}
S.~Weinberg,
Phys. Rev. Lett. \textbf{110}, 261601 (2013)
doi:10.1103/PhysRevLett.110.261601
[arXiv:1303.0342 [hep-ph]].

\bibitem{Knecht:2013yqa}
M.~Knecht and S.~Peris,
Phys. Rev. D \textbf{88}, 036016 (2013)
doi:10.1103/PhysRevD.88.036016
[arXiv:1307.1273 [hep-ph]].

\bibitem{Lebed:2013aka}
R.~F.~Lebed,
Phys. Rev. D \textbf{88}, 057901 (2013)
doi:10.1103/PhysRevD.88.057901
[arXiv:1308.2657 [hep-ph]].


\bibitem{Cohen:2014tga}
T.~D.~Cohen and R.~F.~Lebed,
Phys. Rev. D \textbf{90}, no.1, 016001 (2014)
doi:10.1103/PhysRevD.90.016001
[arXiv:1403.8090 [hep-ph]].



\bibitem{Cohen:2014vta}
T.~Cohen, F.~J.~Llanes-Estrada, J.~R.~Pelaez and J.~Ruiz de Elvira,
Phys. Rev. D \textbf{90}, no.3, 036003 (2014)
doi:10.1103/PhysRevD.90.036003
[arXiv:1405.4831 [hep-ph]].


\bibitem{Maiani:2016hxw}
L.~Maiani, A.~D.~Polosa and V.~Riquer,
JHEP \textbf{06}, 160 (2016)
doi:10.1007/JHEP06(2016)160
[arXiv:1605.04839 [hep-ph]].


\bibitem{Lucha:2017gqq}
W.~Lucha, D.~Melikhov and H.~Sazdjian,
Eur. Phys. J. C \textbf{77}, no.12, 866 (2017)
doi:10.1140/epjc/s10052-017-5437-x
[arXiv:1710.08316 [hep-ph]].


\bibitem{Jaffe:2007id}
R.~L.~Jaffe,
AIP Conf. Proc. \textbf{964}, no.1, 1-13 (2007)
doi:10.1063/1.2823850
[arXiv:hep-ph/0701038 [hep-ph]].

\bibitem{Jaffe:2008zz}
R.~L.~Jaffe,
Nucl. Phys. A \textbf{804}, 25-47 (2008)
doi:10.1016/j.nuclphysa.2008.01.009


\bibitem{Jaffe:1981}
R.~L.~Jaffe, MIT-CTP-951, Rapporteur’s talk presented at Lepton Photon Symp., Bonn, Germany, Aug 24-29, 1981. 


\bibitem{Dashen:1993jt}
R.~F.~Dashen, E.~E.~Jenkins and A.~V.~Manohar,
Phys. Rev. D \textbf{49}, 4713 (1994)
[erratum: Phys. Rev. D \textbf{51}, 2489 (1995)]
doi:10.1103/PhysRevD.51.2489
[arXiv:hep-ph/9310379 [hep-ph]].

\bibitem{Manohar:1998xv}
A.~V.~Manohar,
[arXiv:hep-ph/9802419 [hep-ph]].

\bibitem{Goity:1996hk}
J.~L.~Goity,
Phys. Lett. B \textbf{414}, 140-148 (1997)
doi:10.1016/S0370-2693(97)01154-4
[arXiv:hep-ph/9612252 [hep-ph]].


\bibitem{Lutz:2001yb}
M.~F.~M.~Lutz and E.~E.~Kolomeitsev,
Nucl. Phys. A \textbf{700}, 193-308 (2002)
doi:10.1016/S0375-9474(01)01312-4
[arXiv:nucl-th/0105042 [nucl-th]].

\bibitem{Pelaez:2010er}
J.~R.~Pelaez, J.~Nebreda and G.~Rios,
Prog. Theor. Phys. Suppl. \textbf{186}, 113-123 (2010)
doi:10.1143/PTPS.186.113
[arXiv:1007.3461 [hep-ph]].


\bibitem{Oller:1998zr}
J.~A.~Oller and E.~Oset,
Phys. Rev. D \textbf{60}, 074023 (1999)
doi:10.1103/PhysRevD.60.074023
[arXiv:hep-ph/9809337 [hep-ph]].

\bibitem{Pelaez:2003dy}
J.~R.~Pelaez,
Phys. Rev. Lett. \textbf{92}, 102001 (2004)
doi:10.1103/PhysRevLett.92.102001
[arXiv:hep-ph/0309292 [hep-ph]].

\bibitem{Weinberg:1978kz}
S.~Weinberg,
Physica A \textbf{96}, no.1-2, 327-340 (1979)
doi:10.1016/0378-4371(79)90223-1


\bibitem{Gasser:1983yg}
J.~Gasser and H.~Leutwyler,
Annals Phys. \textbf{158}, 142 (1984)
doi:10.1016/0003-4916(84)90242-2


\bibitem{Gasser:1984gg}
J.~Gasser and H.~Leutwyler,
Nucl. Phys. B \textbf{250}, 465-516 (1985)
doi:10.1016/0550-3213(85)90492-4

\bibitem{Peris:1994dh}
S.~Peris and E.~de Rafael,
Phys. Lett. B \textbf{348}, 539-542 (1995)
doi:10.1016/0370-2693(95)00160-M
[arXiv:hep-ph/9412343 [hep-ph]].


\bibitem{Ledwig:2014cla}
T.~Ledwig, J.~Nieves, A.~Pich, E.~Ruiz Arriola and J.~Ruiz de Elvira,
Phys. Rev. D \textbf{90}, no.11, 114020 (2014)
doi:10.1103/PhysRevD.90.114020
[arXiv:1407.3750 [hep-ph]].


\bibitem{Oller:1997ng}
J.~A.~Oller, E.~Oset and J.~R.~Pelaez,
Phys. Rev. Lett. \textbf{80}, 3452-3455 (1998)
doi:10.1103/PhysRevLett.80.3452
[arXiv:hep-ph/9803242 [hep-ph]].


\bibitem{Oller:1998hw}
J.~A.~Oller, E.~Oset and J.~R.~Pelaez,
Phys. Rev. D \textbf{59}, 074001 (1999)
[erratum: Phys. Rev. D \textbf{60}, 099906 (1999); erratum: Phys. Rev. D \textbf{75}, 099903 (2007)]
doi:10.1103/PhysRevD.59.074001
[arXiv:hep-ph/9804209 [hep-ph]].

\bibitem{GomezNicola:2001as}
A.~Gomez Nicola and J.~R.~Pelaez,
Phys. Rev. D \textbf{65}, 054009 (2002)
doi:10.1103/PhysRevD.65.054009
[arXiv:hep-ph/0109056 [hep-ph]].


\bibitem{Bijnens:2014lea}
J.~Bijnens and G.~Ecker,
Ann. Rev. Nucl. Part. Sci. \textbf{64}, 149-174 (2014)
doi:10.1146/annurev-nucl-102313-025528
[arXiv:1405.6488 [hep-ph]].

\bibitem{Pelaez:2004xp}
J.~R.~Pelaez,
Mod. Phys. Lett. A \textbf{19}, 2879-2894 (2004)
doi:10.1142/S0217732304016160
[arXiv:hep-ph/0411107 [hep-ph]].


\bibitem{Tanabashi:2018oca}
  M.~Tanabashi {\it et al.} [Particle Data Group],
  Phys.\ Rev.\ D {\bf 98}, no. 3, 030001 (2018).
  doi:10.1103/PhysRevD.98.030001

  
\bibitem{Sun:2005uk}
Z.~X.~Sun, L.~Y.~Xiao, Z.~Xiao and H.~Q.~Zheng,
Mod. Phys. Lett. A \textbf{22}, 711-718 (2007)
doi:10.1142/S0217732307023304
[arXiv:hep-ph/0503195 [hep-ph]].

\bibitem{Salas-Bernardez:2020hua}
A.~Salas-Bern\'ardez, F.~J.~Llanes-Estrada, J.~Escudero-Pedrosa and J.~A.~Oller,
[arXiv:2010.13709 [hep-ph]].


\bibitem{Dai:2011bs}
L.~Y.~Dai, X.~G.~Wang and H.~Q.~Zheng,
Commun. Theor. Phys. \textbf{57}, 841-848 (2012)
doi:10.1088/0253-6102/57/5/15
[arXiv:1108.1451 [hep-ph]].


\bibitem{Dai:2012kf}
L.~Y.~Dai, X.~G.~Wang and H.~Q.~Zheng,
Commun. Theor. Phys. \textbf{58}, 410-414 (2012)
doi:10.1088/0253-6102/58/3/15
[arXiv:1206.5481 [hep-ph]].


\bibitem{Morgan:1992ge}
D.~Morgan,
Nucl. Phys. A \textbf{543}, 632-644 (1992)
doi:10.1016/0375-9474(92)90550-4
 
\bibitem{Pelaez:2006nj}
J.~R.~Pelaez and G.~Rios,
Phys. Rev. Lett. \textbf{97}, 242002 (2006)
doi:10.1103/PhysRevLett.97.242002
[arXiv:hep-ph/0610397 [hep-ph]].

\bibitem{Bijnens:1995yn}
J.~Bijnens, G.~Colangelo, G.~Ecker, J.~Gasser and M.~E.~Sainio,
Phys. Lett. B \textbf{374}, 210-216 (1996)
doi:10.1016/0370-2693(96)00165-7
[arXiv:hep-ph/9511397 [hep-ph]].

\bibitem{RuizdeElvira:2010cs}
J.~Ruiz de Elvira, J.~R.~Pelaez, M.~R.~Pennington and D.~J.~Wilson,
Phys. Rev. D \textbf{84}, 096006 (2011)
doi:10.1103/PhysRevD.84.096006
[arXiv:1009.6204 [hep-ph]].

\bibitem{Nieves:2009ez}
J.~Nieves and E.~Ruiz Arriola,
Phys. Rev. D \textbf{80}, 045023 (2009)
doi:10.1103/PhysRevD.80.045023
[arXiv:0904.4344 [hep-ph]].

\bibitem{Omnes:1958hv}
R.~Omnes,
Nuovo Cim. \textbf{8}, 316-326 (1958)
doi:10.1007/BF02747746

\bibitem{Dai:2017uao}
L.~Y.~Dai and U.~G.~Mei\ss{}ner,
Phys. Lett. B \textbf{783}, 294-300 (2018)
doi:10.1016/j.physletb.2018.06.071
[arXiv:1706.10123 [hep-ph]].

\bibitem{Dai:2018fmx}
L.~Y.~Dai, X.~W.~Kang and U.~G.~Mei\ss{}ner,
Phys. Rev. D \textbf{98}, no.7, 074033 (2018)
doi:10.1103/PhysRevD.98.074033
[arXiv:1808.05057 [hep-ph]].

\bibitem{Guo:2012ym}
Z.~H.~Guo, J.~A.~Oller and J.~Ruiz de Elvira,
Phys. Lett. B \textbf{712}, 407-412 (2012)
doi:10.1016/j.physletb.2012.05.021
[arXiv:1203.4381 [hep-ph]].

\bibitem{Guo:2012yt}
Z.~H.~Guo, J.~A.~Oller and J.~Ruiz de Elvira,
Phys. Rev. D \textbf{86}, 054006 (2012)
doi:10.1103/PhysRevD.86.054006
[arXiv:1206.4163 [hep-ph]].


\bibitem{Wolkanowski:2015jtc}
T.~Wolkanowski, M.~So\l{}tysiak and F.~Giacosa,
Nucl. Phys. B \textbf{909}, 418-428 (2016)
doi:10.1016/j.nuclphysb.2016.05.025
[arXiv:1512.01071 [hep-ph]].

\bibitem{Tornqvist:1982yv}
N.~A.~Tornqvist,
Phys. Rev. Lett. \textbf{49}, 624-627 (1982)
doi:10.1103/PhysRevLett.49.624


\bibitem{Achasov:1994iu}
N.~N.~Achasov and G.~N.~Shestakov,
Phys. Rev. D \textbf{49}, 5779-5784 (1994)
doi:10.1103/PhysRevD.49.5779


\bibitem{vanBeveren:2006ua}
E.~van Beveren, D.~V.~Bugg, F.~Kleefeld and G.~Rupp,
Phys. Lett. B \textbf{641}, 265-271 (2006)
doi:10.1016/j.physletb.2006.08.051
[arXiv:hep-ph/0606022 [hep-ph]].


\bibitem{Giacosa:2006tf}
F.~Giacosa,
Phys. Rev. D \textbf{75}, 054007 (2007)
doi:10.1103/PhysRevD.75.054007
[arXiv:hep-ph/0611388 [hep-ph]].

\bibitem{Zhou:2010ra}
Z.~Y.~Zhou and Z.~Xiao,
Phys. Rev. D \textbf{83}, 014010 (2011)
doi:10.1103/PhysRevD.83.014010
[arXiv:1007.2072 [hep-ph]].

\bibitem{Lukashov:2019dir}
M.~S.~Lukashov and Y.~A.~Simonov,
Phys. Rev. D \textbf{101}, no.9, 094028 (2020)
doi:10.1103/PhysRevD.101.094028
[arXiv:1909.10384 [hep-ph]].

\bibitem{Geng:2008ag}
L.~S.~Geng, E.~Oset, J.~R.~Pelaez and L.~Roca,
Eur. Phys. J. A \textbf{39}, 81-87 (2009)
doi:10.1140/epja/i2008-10689-y
[arXiv:0811.1941 [hep-ph]].

\bibitem{Sannino:1995ik}
F.~Sannino and J.~Schechter,
Phys. Rev. D \textbf{52}, 96-107 (1995)
doi:10.1103/PhysRevD.52.96
[arXiv:hep-ph/9501417 [hep-ph]].



\bibitem{Nieves:2011gb}
J.~Nieves, A.~Pich and E.~Ruiz Arriola,
Phys. Rev. D \textbf{84}, 096002 (2011)
doi:10.1103/PhysRevD.84.096002
[arXiv:1107.3247 [hep-ph]].

\bibitem{Nieves:2009kh}
J.~Nieves and E.~Ruiz Arriola,
Phys. Lett. B \textbf{679}, 449-453 (2009)
doi:10.1016/j.physletb.2009.08.021
[arXiv:0904.4590 [hep-ph]].


\bibitem{Lucini:2012gg}
B.~Lucini and M.~Panero,
Phys. Rept. \textbf{526}, 93-163 (2013)
doi:10.1016/j.physrep.2013.01.001
[arXiv:1210.4997 [hep-th]]; 

\bibitem{Lucini:2013qja}
B.~Lucini and M.~Panero,
Prog. Part. Nucl. Phys. \textbf{75}, 1-40 (2014)
doi:10.1016/j.ppnp.2014.01.001
[arXiv:1309.3638 [hep-th]].


\bibitem{DelDebbio:2007wk}
L.~Del Debbio, B.~Lucini, A.~Patella and C.~Pica,
JHEP \textbf{03}, 062 (2008)
doi:10.1088/1126-6708/2008/03/062
[arXiv:0712.3036 [hep-th]].

\bibitem{Hietanen:2009tu}
A.~Hietanen, R.~Narayanan, R.~Patel and C.~Prays,
Phys. Lett. B \textbf{674}, 80-82 (2009)
doi:10.1016/j.physletb.2009.02.054
[arXiv:0901.3752 [hep-lat]].


\bibitem{Bali:2013kia}
G.~S.~Bali, F.~Bursa, L.~Castagnini, S.~Collins, L.~Del Debbio, B.~Lucini and M.~Panero,
JHEP \textbf{06}, 071 (2013)
doi:10.1007/JHEP06(2013)071
[arXiv:1304.4437 [hep-lat]].


\bibitem{DeGrand:2016pur}
T.~DeGrand and Y.~Liu,
Phys. Rev. D \textbf{94}, no.3, 034506 (2016)
[erratum: Phys. Rev. D \textbf{95}, no.1, 019902 (2017)]
doi:10.1103/PhysRevD.94.034506
[arXiv:1606.01277 [hep-lat]].

\bibitem{Perez:2020fqn}
M.~G.~P\'erez, A.~Gonz\'alez-Arroyo and M.~Okawa,
PoS \textbf{LATTICE2019}, 113 (2019)
doi:10.22323/1.363.0113
[arXiv:2001.00172 [hep-lat]].


\bibitem{Hernandez:2019qed}
P.~Hern\'andez, C.~Pena and F.~Romero-L\'opez,
Eur. Phys. J. C \textbf{79}, no.10, 865 (2019)
doi:10.1140/epjc/s10052-019-7395-y
[arXiv:1907.11511 [hep-lat]].

\bibitem{Donini:2020qfu}
A.~Donini, P.~Hern\'andez, C.~Pena and F.~Romero-L\'opez,
Eur. Phys. J. C \textbf{80}, no.7, 638 (2020)
doi:10.1140/epjc/s10052-020-8192-3
[arXiv:2003.10293 [hep-lat]].

\bibitem{Nebreda:2011cp}
J.~Nebreda, J.~R.~Pelaez and G.~Rios,
Phys. Rev. D \textbf{84}, 074003 (2011)
doi:10.1103/PhysRevD.84.074003
[arXiv:1107.4200 [hep-ph]].




\bibitem{ua1anomaly}
S.~L.~Adler and W.~A.~Bardeen,
Phys. Rev. \textbf{182}, 1517-1536 (1969)
doi:10.1103/PhysRev.182.1517

W.~A.~Bardeen,
Phys. Rev. \textbf{184}, 1848-1857 (1969)
doi:10.1103/PhysRev.184.1848


K.~Fujikawa,
Phys. Rev. D \textbf{21}, 2848 (1980)
doi:10.1103/PhysRevD.21.2848


\bibitem{Kaiser:2000gs}
  R.~Kaiser and H.~Leutwyler,
  Eur.\ Phys.\ J.\ C {\bf 17}, 623 (2000)
  doi:10.1007/s100520000499
  [hep-ph/0007101].

\bibitem{ua1nc}

E.~Witten,
Nucl. Phys. B \textbf{156}, 269-283 (1979)
doi:10.1016/0550-3213(79)90031-2;

S.~R.~Coleman and E.~Witten,
Phys. Rev. Lett. \textbf{45}, 100 (1980)
doi:10.1103/PhysRevLett.45.100;

G.~Veneziano,
Nucl. Phys. B \textbf{159}, 213-224 (1979)
doi:10.1016/0550-3213(79)90332-8


\bibitem{tHooft:1976rip}
G.~'t Hooft,
Phys. Rev. Lett. \textbf{37}, 8-11 (1976)
doi:10.1103/PhysRevLett.37.8


\bibitem{Witten:1978bc}
E.~Witten,
Nucl. Phys. B \textbf{149}, 285-320 (1979)
doi:10.1016/0550-3213(79)90243-8


\bibitem{Schafer:2002af}
T.~Sch\"afer,
Phys. Rev. D \textbf{66}, 076009 (2002)
doi:10.1103/PhysRevD.66.076009
[arXiv:hep-ph/0206062 [hep-ph]].
  
  
\bibitem{Guo:2011pa}
Z.~H.~Guo and J.~A.~Oller,
Phys. Rev. D \textbf{84}, 034005 (2011)
doi:10.1103/PhysRevD.84.034005
[arXiv:1104.2849 [hep-ph]].

 

\bibitem{Guo:2016zep}
Z.~H.~Guo, L.~Liu, U.~G.~Mei\ss{}ner, J.~A.~Oller and A.~Rusetsky,
Phys. Rev. D \textbf{95}, no.5, 054004 (2017)
doi:10.1103/PhysRevD.95.054004
[arXiv:1609.08096 [hep-ph]].

\bibitem{Gao:2019idb}
R.~Gao, Z.~H.~Guo and J.~Y.~Pang,
Phys. Rev. D \textbf{100}, no.11, 114028 (2019)
doi:10.1103/PhysRevD.100.114028
[arXiv:1907.01787 [hep-ph]].

\bibitem{HerreraSiklody:1996pm}
  P.~Herrera-Siklody, J.~I.~Latorre, P.~Pascual and J.~Taron,
  Nucl.\ Phys.\ B {\bf 497}, 345 (1997)
  doi:10.1016/S0550-3213(97)00260-5
  [hep-ph/9610549].

\bibitem{Gu:2018swy}
X.~W.~Gu, C.~G.~Duan and Z.~H.~Guo,
Phys. Rev. D \textbf{98}, no.3, 034007 (2018)
doi:10.1103/PhysRevD.98.034007
[arXiv:1803.07284 [hep-ph]].

\bibitem{Guo:2015xva}
X.~K.~Guo, Z.~H.~Guo, J.~A.~Oller and J.~J.~Sanz-Cillero,
JHEP \textbf{06}, 175 (2015)
doi:10.1007/JHEP06(2015)175
[arXiv:1503.02248 [hep-ph]].

\bibitem{Oller:2000fj}
J.~A.~Oller and U.~G.~Meissner,
Phys. Lett. B \textbf{500}, 263-272 (2001)
doi:10.1016/S0370-2693(01)00078-8
[arXiv:hep-ph/0011146 [hep-ph]].

\bibitem{Oller:1999me}
J.~A.~Oller,
Phys. Lett. B \textbf{477}, 187-194 (2000)
doi:10.1016/S0370-2693(00)00185-4
[arXiv:hep-ph/9908493 [hep-ph]].

\bibitem{Lacour:2009ej}
A.~Lacour, J.~A.~Oller and U.~G.~Meissner,
Annals Phys. \textbf{326}, 241-306 (2011)
doi:10.1016/j.aop.2010.06.012
[arXiv:0906.2349 [nucl-th]].

\bibitem{Ecker:1988te}
G.~Ecker, J.~Gasser, A.~Pich and E.~de Rafael,
Nucl. Phys. B \textbf{321}, 311-342 (1989)
doi:10.1016/0550-3213(89)90346-5

\bibitem{Meissner:2000bc}
U.~G.~Meissner and J.~A.~Oller,
Nucl. Phys. A \textbf{679}, 671-697 (2001)
doi:10.1016/S0375-9474(00)00367-5
[arXiv:hep-ph/0005253 [hep-ph]].

\bibitem{Oller:2000ug}
J.~A.~Oller, E.~Oset and J.~E.~Palomar,
Phys. Rev. D \textbf{63}, 114009 (2001)
doi:10.1103/PhysRevD.63.114009
[arXiv:hep-ph/0011096 [hep-ph]].

\bibitem{Albaladejo:2010tj}
M.~Albaladejo, J.~A.~Oller and L.~Roca,
Phys. Rev. D \textbf{82}, 094019 (2010)
doi:10.1103/PhysRevD.82.094019
[arXiv:1011.1434 [hep-ph]].


\bibitem{Briceno:2016mjc}
R.~A.~Briceno, J.~J.~Dudek, R.~G.~Edwards and D.~J.~Wilson,
Phys. Rev. Lett. \textbf{118}, no.2, 022002 (2017)
doi:10.1103/PhysRevLett.118.022002
[arXiv:1607.05900 [hep-ph]].

\bibitem{Briceno:2017qmb}
R.~A.~Briceno, J.~J.~Dudek, R.~G.~Edwards and D.~J.~Wilson,
Phys. Rev. D \textbf{97}, no.5, 054513 (2018)
doi:10.1103/PhysRevD.97.054513
[arXiv:1708.06667 [hep-lat]].

\end{thebibliography}
\end{document}